\documentclass[12pt] {elsart} 

\usepackage{amsmath}

\usepackage{placeins}

\usepackage{amssymb}

\usepackage{graphicx}

\begin{document}

\begin{frontmatter}

\title{Quasi Equilibrium Grid Algorithm: geometric construction for model reduction}

\author{Eliodoro Chiavazzo},
\ead{chiavazzo@lav.mavt.ethz.ch}
\author{Iliya V.\ Karlin}
\ead{karlin@lav.mavt.ethz.ch}
\address{Aerothermochemistry and
Combustion Systems Laboratory (LAV), ETHZ CH-8092 Zurich, Switzerland.}

\begin{abstract}
The Method of Invariant Grid (MIG) is an iterative procedure for model reduction in chemical kinetics which is based on the notion of Slow Invariant Manifold (SIM) \cite{1}-\cite{book}. Important role, in that method, is played by the initial grid which, once refined, gives a description of the invariant manifold: the invariant grid. A convenient way to get a first approximation of the SIM is given by the Spectral Quasi Equilibrium Manifold (SQEM) \cite{1}-\cite{ChGoKa07}. In the present paper, a flexible numerical method to construct the discrete analog of a Quasi Equilibrium Manifold, in any dimension, is presented. That object is named Quasi Equilibrium Grid (QEG), while the procedure Quasi Equilibrium Grid Algorithm. Extensions of the QEM notion are also suggested. The QEG is a numerical tool which can be used to find a grid-based approximation for the locus of minima of a convex function under some linear constraints. The method is validated by construction of one and two-dimensional grids for model hydrogen oxidation reaction.      
\end{abstract}


\end{frontmatter}

\section{Introduction}
Relaxation of complex systems is often characterized by a fast dynamics during a short initial stage, while the remaining period lasts much longer and it evolves along low-dimensional surfaces in the phase space known as Slow Invariant Manifolds (SIM). In that scenario, a simplified macroscopic description of a complex system can be attained by extracting only the slow dynamics and neglecting the fast one. For this reason, much effort was spent to develop model reduction methods (the Method of Invariant Grid (MIG) \cite{1,ChGoKa07,2,book}, the Intrinsic Low Dimensional Manifold method (ILDM) \cite{3,4}, the Computational Singular Perturbation method (CSP) \cite{LG91,5,6}, etc.) based on the notion of SIM. The introduction of a convex Lyapunov function $G$, whenever the complex system is supported by such a function, also proves to be very helpful in model reduction \cite{book,GK92}. Indeed, it was shown that, through a $G$ function, good approximations of the SIM can be found (e.g. by constructing the Spectral Quasi Equilibrium Manifold - SQEM - or the Symmetric Entropic Intrinsic Low Dimensional Manifold - SEILDM - \cite{1,ChGoKa07}) and refined by some efficient MIG iterations. Moreover, it has been shown that the notion of QEM is also very useful in different fields. For example, it was used in the implementation of Lattice Boltzmann schemes \cite{KFO99,AMAKFB99}. Construction of a QEM is analytically possible by using the Lagrange multipliers method. However, its implementation becomes too complicated as soon as the number of variables of the problem becomes large: efficient methods, for constructing large dimensional QEM, are still missing. Therefore, in the present paper the notion of Quasi Equilibrium Grid (QEG) will be introduced, as a discrete analog of QEM, and a constructive algorithm, applicable in any dimension, will be developed. The procedure suggested proves to be a very flexible tool, so it is possible to get some other SIM approximations, all based on the previous algorithm, even more accurate than the QEG itself. 
\section{Paper organization}
The paper is organized as follows. In Section \ref{basics}, some basic notions are outlined: in particular, the general equations of dissipative reaction kinetics are reviewed, in the notations which are used throughout the paper. At the end of that Section, the Method of Invariant Grid (MIG) and the notion of thermodynamic projector are briefly discussed (Section \ref{MIG}). In Section \ref{QEM_intr}, the QEM definition and its geometrical interpretation is given, while in Section \ref{main_alg} the 1D Quasi Equilibrium Grid Algorithm is presented. That algorithm is also illustrated, by means of an example, in Section \ref{ex_1D}. The 1D Algorithm extension to multi-dimensional grids is developed in Section \ref{Ext_multi}. In particular, two possible extension strategies are analyzed: the \textit{straightforward extension} (Section \ref{stra_exte}) and, by following the general idea given in \cite{2}, the \textit{flag extension} (Section \ref{flag_ext}). Here, it is also shown how the flexibility of the flag extension allows to get SIM approximation which is better than the Spectral-QEG (Section \ref{Beyond}): the notions of \textit{Guided-QEG} and \textit{Symmetric Entropic Guided-QEG} are introduced. An illustrative example, in Section \ref{2D_example}, shows how those different extension techniques work in practice. In order to find out how accurate is their SIM description, they are also compared on the base of the invariance defect (Section \ref{compar_2D}). Finally, results are discussed in Section \ref{conclus}.

\section{Theoretical background} \label{basics}

\subsection{Dissipative reaction kinetics}\label{Diss_kin}
In a closed system with \emph{n} chemical species $A_1,...,A_n$,
participating in a complex reaction, a generic reversible reaction step can
be written as a stoichiometric equation:
\begin{equation}\label{a}
\alpha _{s1} A_1  + ... + \alpha _{sn} A_n  \rightleftharpoons
\beta _{s1} A_1  + ... + \beta _{sn} A_n,
\end{equation}
where \emph{s} is the reaction index, $s=1,...,r$ (\emph{r} steps
in total), and the integers $\alpha_{si}$ and $\beta_{si}$ are
stoichiometric coefficients of the step \emph{s}. For each
reaction step, we can introduce \emph{n}-component vectors
$\mbox{\boldmath$\alpha$}_s$ and $\mbox{\boldmath$\beta$}_s$,
with components $\alpha_{si}$ and $\beta_{si}$, and the
stoichiometric vector
$\mbox{\boldmath$\gamma$}_s$=$\mbox{\boldmath$\beta$}_s$-$\mbox{\boldmath$\alpha$}_s$.
For every $A_i$ the \emph{extensive variable} $N_i$ describes the
number of particles of that species. If \emph{V} is the volume,
then the concentration of $A_i$ is $c_i=N_i/V$. Dynamics of the
species concentration according to the stoichiometric mechanism (\ref{a}) reads:
\begin{equation}\label{b}
\begin{array}{*{20}c}
   {\dot {\emph{\textbf{N}}} = V\emph{\textbf{J}}(\emph{\textbf{c}}),} & {} & {\emph{\textbf{J}}(\emph{\textbf{c}}) = \sum\nolimits_{s = 1}^r {\mbox{\boldmath$\gamma$}_s W_s (\emph{\textbf{c}}),} }  \\
\end{array}
\end{equation}
where dot denotes the time derivative and $W_s(\emph{\textbf{c}})$
is the reaction rate function of the step \emph{s}. In particular,
the polynomial form of the reaction rate function is provided by
\emph{the mass action law}:
\begin{equation}\label{c}
W_s (\emph{\textbf{c}}) = W_s^ +  (\emph{\textbf{c}}) - W_s^ -
(\emph{\textbf{c}}) = k_s^ + (T)\prod\limits_{i = 1}^n
{c_i^{\alpha _i } - } k_s^ - (T)\prod\limits_{i = 1}^n {c_i^{\beta
_i } ,}
\end{equation}
where $k_s^+(T)$ and $k_s^-(T)$ are the constants
of the direct and of the inverse reactions rates of the step
\emph{s} respectively. The most popular form of their dependence is given
by the Arrhenius equation:
\begin{equation}
k_s^ \pm  (T) = a_s^ \pm  T^{b_s^ \pm  } \exp (S_s^ \pm  /k_B
)\exp ( - H_s^ \pm  /k_B T). \nonumber
\end{equation}
In the latter equation, $a_s^ \pm$,$b_s^ \pm$ are constants and $H_s^
\pm$, $S_s^ \pm$ activation enthalpies and entropies respectively.
The rate constants are not independent. Indeed, the
\emph{principle of detail balance} gives a relation between these
quantities:
\begin{equation}\label{detail.balance}
\begin{array}{*{20}c}
   {W_s^ +  (\emph{\textbf{c}}^{eq} ) = W_s^ -  (\emph{\textbf{c}}^{eq} ),} & {} & {\forall s = 1,...,r,}  \\
\end{array}
\end{equation}
where the positive vector $\emph{\textbf{c}}^{eq}(T)$ is the
equilibrium of the system (\ref{b}). In order to obtain a closed
system of equations, one should supply an equation for the volume
\emph{V}. For an isolated system the extra-equations are
$\emph{U},\emph{V}=\emph{const}$ (where \emph{U} is the internal
energy), for an isochoric isothermal system we get
$\emph{V},\emph{T}=\emph{const}$, and so forth. For example,
equation (\ref{b}) in the latter case simply takes the form:
\begin{equation}\label{d}
    \dot{\emph{\textbf{c}}}=\sum\limits_{s = 1}^r {\mbox{\boldmath$\gamma$}_s } W_s
    (\emph{\textbf{c}})=\emph{\textbf{J}}(\emph{\textbf{c}}).
\end{equation}
Finally, also other linear constraints, related to the conservation of atoms, must be considered.
In general such conservation laws can have the following form:
\begin{equation}\label{cons.law}
\emph{\textbf{D}}\emph{\textbf{c}} = \emph{\textbf{const}},
\end{equation}
where $l$ fixed and linearly independent vectors
$\mbox{\boldmath$d$}_i$ are the rows of the $l \times n$ matrix $\emph{\textbf{D}}$, and $\emph{\textbf{const}}$ is a constant vector. 

\subsection{Outline of the method of invariant grid}\label{MIG}
In this section, we give an outline of the MIG for chemical kinetics. For details see Refs.
\cite{1,ChGoKa07,2,book,GK92}.

\subsubsection{Thermodynamic potential}

If we turn our attention to perfectly stirred closed chemically
active mixtures, then dissipative properties of such systems can
be characterized with a thermodynamic potential which is the
Lyapunov function of equation (\ref{b}). That function implements Second Law
of thermodynamics: it means that during the concentrations evolution in time,
from the initial condition to the equilibrium state, the Lyapunov function must
decrease monotonically. Therefore if $G(\emph{\textbf{c}})$ is
the Lyapunov function, $\emph{\textbf{c}}^{eq}$ (equilibrium
state) is its point of global minimum in the phase space. A simple
example of a function $G$ is given by the free energy of ideal gas in a
constant volume and under a constant temperature:
\begin{equation}\label{Gfunc}
G = \sum\limits_{i = 1}^n {c_i [\ln (c_i /c_i^{eq} ) - 1]}.
\end{equation}
When $G$ is known, also its gradient $\nabla G$ and the matrix of second
derivatives $\mbox{\boldmath$H$}=\parallel\partial ^2
G/\partial c_i \partial c_j\parallel$ can be evaluated, so that it
is possible to introduce the thermodynamic scalar product as
follows:
\begin{equation}\label{entr.prod}
    \left\langle \mbox{\boldmath$x$},\mbox{\boldmath$y$} \right\rangle=(\mbox{\boldmath$x$},\mbox{\boldmath$H$}\mbox{\boldmath$y$}),
\end{equation}
where the notation $(,)$ is the usual Euclidean scalar product.

\subsubsection{The invariance condition}
Let us consider $\mbox{\boldmath$\Omega$}$ as a manifold of a
reduced description. The invariance requirement reads:
\begin{equation}\label{inv.cond}
    \begin{array}{*{20}c}
    {\mbox{\boldmath$c$}(0)\in\mbox{\boldmath$\Omega$}\Rightarrow\mbox{\boldmath$c$}(t)\in\mbox{\boldmath$\Omega$},} & {} & {\forall
t \ge 0.}
  \end{array}
\end{equation}
Let $\mbox{\boldmath$P$}$ be a projector on the tangent bundle of
the manifold \mbox{\boldmath$\Omega$}. The manifold
$\mbox{\boldmath$\Omega$}$ is invariant with respect to the system
(\ref{b}) if and only if the following invariance
equation (IE) holds:
\begin{equation}\label{inv.equation}
 \begin{array}{*{20}c}
    {[1-\mbox{\boldmath$P$}]\mbox{\boldmath$J$}(\mbox{\boldmath$c$})=0,} & {} & {\forall \mbox{\boldmath$c$} \in \mbox{\boldmath$\Omega$}.}
 \end{array}
\end{equation}
When the manifold is not invariant, it is not able
to satisfy the invariance condition so that:
\begin{equation}\label{inv.defect}
    \exists \mbox{\boldmath$c$}_0 \\:\\
    \Delta_0=[1-\mbox{\boldmath$P$}]\mbox{\boldmath$J$}(\mbox{\boldmath$c$}_0)\neq0,
\end{equation}
where $\Delta_0$ is the defect of invariance. One way to
find the SIM is to solve the IE iteratively starting from
an appropriate initial manifold.

\subsubsection{Thermodynamic projector} \label{SecTherPro}
Let us now discuss further the projector appearing in the invariance
equation. It is an operator which for each point
$\mbox{\boldmath$c$}\in\mbox{\boldmath$\Omega$}$ projects the vectors
$\mbox{\boldmath$J$}(\mbox{\boldmath$c$})$ onto the tangent subspace
of the manifold producing, in this way, the induced vector field
$\mbox{\boldmath$P$}\mbox{\boldmath$J$}(\mbox{\boldmath$c$})$. In general,
condition (\ref{inv.equation}) does not require any special constraint for
the projector $\mbox{\boldmath$P$}$. However, the thermodynamic properties of the kinetic
equations (\ref{b}) define the projector unambiguously \cite{1,book,GK92}.
To this end, let us define a differential of $G$, that is linear
functional:
\begin{equation}\label{thermo.functional}
DG(\mbox{\boldmath$x$}) = (\nabla
G(\mbox{\boldmath$c$}),\mbox{\boldmath$x$}).
\end{equation}
A special class of projectors is the
thermodynamic one. If a projector belongs to this class then the
induced vector field respects the dissipation inequality:
\begin{equation}\label{thermo.condition.1}
    \begin{array}{*{20}c}
    {DG (\mbox{\boldmath$P$}\mbox{\boldmath$J$})\leq0,} & {} & {\forall\mbox{\boldmath$c$}\in\mbox{\boldmath$\Omega$}.}
    \end{array}  
\end{equation}
It has been shown that a projector $\mbox{\boldmath$P$}$ respects the (\ref{thermo.condition.1}) if and only if \cite{GK92}:
\begin{equation}\label{thermo.condition.2}
\begin{array}{*{20}c}
    {\ker\mbox{\boldmath$P$}\subseteq \ker DG,} & {} & {\forall \mbox{\boldmath$c$} \in \mbox{\boldmath$\Omega$},}
\end{array}
\end{equation}
where $\ker$ denotes the null-space of an operator. It is clear
now that if one wants to solve equation (\ref{inv.equation}), then
a projector must be specified. Here we remind the way to
construct the thermodynamic projector which will be used in MIG
procedure \cite{1}. This projector depends on the concentration point
\mbox{\boldmath$c$} and on the tangent space to the manifold
$\mbox{\boldmath$\Omega$}$.

We are looking for a grid approximation of a $q$-dimensional SIM.
Let $\mathcal{G}$ be a discrete subset of a $q$-dimensional
parameter space $\mbox{\boldmath$R$}^q$ and let
$F\mid_{\mathcal{G}}$ be a mapping of $\mathcal{G}$ into the
concentration space. If we select an approximation procedure to
restore the smooth map $F$ from the discrete map
$F\mid_{\mathcal{G}}$ (we need  a very small part of $F$,
derivatives of $F$ in the grid points only), then the derivatives
$\mbox{\boldmath$f$}_i=\partial F/\partial y_i$ are available, and
for each grid point the tangent space is:
\begin{equation}\label{tangent.space}
\begin{array}{*{20}c}
{T_y=Lin\{\mbox{\boldmath$f$}_i\},} & {} & {i=1,...,n.}
\end{array}
\end{equation}
We assume that one of points $y \in \mathcal{G}$ maps into the
equilibrium, and in other points intersection of the manifold with $G$
levels is transversal (i.e. $(DG)_{F(y)}(x) \neq  0$ for some $x
\in T_y$). Let us consider the subspace $T_{0y}=(T_y \cap \ker DG)$. In order to define the thermodynamic
projector, it is required, if $T_{0y} \neq T_y$, to introduce the
vector $\mbox{\boldmath$e$}_y$ which satisfies the following conditions:
\begin{equation}\nonumber
\left\{ \begin{array}{l}
\mbox{\boldmath$e$}_y  \in T_y , \\
\left\langle {\mbox{\boldmath$e$}_y,\mbox{\boldmath$x$}} \right\rangle=0, \forall \mbox{\boldmath$x$} \in T_{0y},\\
DG (\mbox{\boldmath$e$}_y ) = 1. \\
 \end{array} \right.
\end{equation}
Let $\mbox{\boldmath$P$}_{0}$ be the orthogonal projector on
$T_{0y}$ with respect to the entropic scalar product
(\ref{entr.prod}), then the thermodynamic projection of a vector $\mbox{\boldmath$x$}$ is
defined as:
\begin{equation}\label{thermo.projector}
\left\{ \begin{array}{l}
 T_{0y}  \ne T_y  \Rightarrow \mbox{\boldmath$P$}   \mbox{\boldmath$x$}  = \mbox{\boldmath$P$}_{0}  \mbox{\boldmath$x$}  + \mbox{\boldmath$e$}_y DG ( \mbox{\boldmath$x$} ) \\
 T_{0y}  = T_y  \Rightarrow \mbox{\boldmath$P$}   \mbox{\boldmath$x$}  = \mbox{\boldmath$P$}_{0} \mbox{\boldmath$x$} . \\
 \end{array} \right.
\end{equation}

\subsubsection{Iterative procedures: the Newton method with incomplete linearization}
When MIG method is applied, not a manifold is searched as a
solution, but a set of concentration points whose defect of
invariance is sufficiently small: let $\mbox{\boldmath$\Omega$}$
denote that solution (invariant grid). MIG is an iterative
procedure: this means that, at the beginning, only an initial
approximation $\mbox{\boldmath$\Omega$}_0$ of
$\mbox{\boldmath$\Omega$}$ is available. In general,
$\mbox{\boldmath$\Omega$}_0$ does not respect the invariance
condition (\ref{inv.equation}) satisfactorily so the
(\ref{inv.defect}) holds: for this reason the position of
$\mbox{\boldmath$c$}_0 \in \mbox{\boldmath$\Omega$}_0$ must be
changed. We can think to correct its position and get a new point ($\mbox{\boldmath$c$}_0 + \delta\mbox{\boldmath$c$}$) with a lower defect of invariance $\mbox{\boldmath$\Delta$}=[1-\mbox{\boldmath$P$}]\mbox{\boldmath$J$}(\mbox{\boldmath$c$}_0 + \delta\mbox{\boldmath$c$})$. If the initial node is ``not far'' from the invariant manifold, a reasonable way to get the node correction $\delta\mbox{\boldmath$c$}$ is to solve the linearized invariance equation where the vector field $\mbox{\boldmath$J$}$ is expanded to the first order and the projector $\mbox{\boldmath$P$}$ to the zeroth order:
\begin{equation}\label{lin_inv_eq}
[1 - \mbox{\boldmath$P$}(\mbox{\boldmath$c$})][\mbox{\boldmath$J$}(\mbox{\boldmath$c$}) + \mbox{\boldmath$L$}(\mbox{\boldmath$c$})\delta \mbox{\boldmath$c$}] = 0.
\end{equation}
$\mbox{\boldmath$L$}$ is the matrix of first derivatives of $\mbox{\boldmath$J$}$ (Jacobian matrix).
The Newton method with incomplete linearization consists of the equation (\ref{lin_inv_eq}) supplied by the extra condition \cite{GK92}:
\begin{equation}\label{ker_proj}
	\mbox{\boldmath$P$}\delta\mbox{\boldmath$c$}=0.
\end{equation}
The additional condition (\ref{ker_proj}) and the atoms balances (\ref{cons.law}) automatically can be taken into account
choosing a basis \{$\mbox{\boldmath$b$}_i$\} in the subspace $\mbox{\boldmath$S$}=(\ker \mbox{\boldmath$P$} \cap \ker \mbox{\boldmath$D$})$.
Let $h = dim(\mbox{\boldmath$S$})$, then the correction can be cast in the form $\delta \mbox{\boldmath$c$} = \sum\nolimits_{i = 1}^h {\delta _i \mbox{\boldmath$b$}_i }$, so that the linearized invariance equation (\ref{lin_inv_eq}) becomes the linear algebraic system in terms of $\delta_i$:
\begin{equation}\label{NM_equation}
\begin{array}{*{20}c}
   {\sum\nolimits_{i = 1}^h {\delta _i \left( {(1 - \mbox{\boldmath$P$})\mbox{\boldmath$L$}\mbox{\boldmath$b$}_i ,\mbox{\boldmath$b$}_k } \right) =  - \left( {(1 - \mbox{\boldmath$P$})\mbox{\boldmath$J$},\mbox{\boldmath$b$}_k } \right)} ,} & {} & {k = 1,...,h.}  \\
\end{array}
\end{equation}
{\it Remark.} Here the usual scalar product $(,)$ was used to get the components of the left-hand side of (\ref{lin_inv_eq}) in the basis vectors \{$\mbox{\boldmath$b$}_i$\}. Nevertheless, a different scalar product can be also used without a loss of generality. 

In the case of the thermodynamic projector, it proves convenient to choose the basis \{$\mbox{\boldmath$b$}_i$\} orthonormal with respect to the entropic scalar product (\ref{entr.prod}) and write the (\ref{NM_equation}) as: 
\begin{equation}\label{NM_equation_entr}
\begin{array}{*{20}c}
   {\sum\nolimits_{i = 1}^h {\delta _i \left\langle {(1 - \mbox{\boldmath$P$})\mbox{\boldmath$L$}\mbox{\boldmath$b$}_i ,\mbox{\boldmath$b$}_k } \right\rangle  =  - \left\langle {(1 - \mbox{\boldmath$P$})\mbox{\boldmath$J$},\mbox{\boldmath$b$}_k } \right\rangle ,} } & {} & {k = 1,...,h.}  \\
\end{array} 
\end{equation}
The projector (\ref{thermo.projector}) is ``almost'' $\left\langle , \right\rangle-$orthogonal ($\left\langle { im \mbox{\boldmath$P$} , \ker \mbox{\boldmath$P$}} \right\rangle   \cong  0$) close to the SIM. Because of that special feature, equation (\ref{NM_equation_entr}) can be approximated and simplified as follows: 
\begin{equation}\label{new.meth.orth}
\begin{array}{*{20}c}
   {\sum\nolimits_{i = 1}^h {\delta _i \left\langle {\mbox{\boldmath$L$} \mbox{\boldmath$b$}_i ,\mbox{\boldmath$b$}_k } \right\rangle  =  - \left\langle {\mbox{\boldmath$J$},\mbox{\boldmath$b$}_k } \right\rangle ,} } & {} & {k = 1,...,h.}  \\
\end{array}
\end{equation}
Note that, in general, an approximation carried out by eq. (\ref{new.meth.orth}) leaves a residual defect (\ref{inv.defect}) in the grid nodes which cannot be completely annihilated. Therefore, when a higher accuracy in the SIM description is required, equation (\ref{NM_equation}) is recommended.

\subsubsection{Iterative procedures: the relaxation method}
An alternative approach to solve eq. (\ref{lin_inv_eq}) is the \emph{relaxation
method}. According to that method the correction is written as
$\mbox{\boldmath$c$} = \mbox{\boldmath$c$}_0 + \tau
(\mbox{\boldmath$c$})\Delta (\mbox{\boldmath$c$})$, and the
quantity $\tau (\mbox{\boldmath$c$})$ is obtained from the condition:
\begin{equation}\nonumber
\left\langle {\Delta ,[1 - \mbox{\boldmath$P$}
][\mbox{\boldmath$J$} + \tau
(\mbox{\boldmath$c$})\mbox{\boldmath$L$} \Delta ]} \right\rangle
 = 0,
\end{equation}
and solving with respect to $\tau$:
\begin{equation}\label{tau.relaxation}
\tau (\mbox{\boldmath$c$}) =  - \frac{{\left\langle {\Delta
,\Delta } \right\rangle  }}{{\left\langle {\Delta
,\mbox{\boldmath$L$} \Delta } \right\rangle }}.
\end{equation}
Equation (\ref{tau.relaxation}) shows that the
relaxation method is explicit, but as it adjusts the node position
acting only along the direction of the defect $\Delta$, typically we expect it to be
less efficient in comparison with the Newton method. On the
other hand, this method is particularly easy to implement.
\section{The initial approximation. The Quasi Equilibrium Manifold}\label{QEM_intr}
Any iterative procedure needs to be supplied by a first approximation. Since it plays an important role for both the convergence and efficiency, that approximation must be chosen very carefully. It was shown that a reasonable way, for initializing the MIG, is to construct the Quasi Equilibrium Manifold (QEM) \cite{1,ChGoKa07}. 
\subsection{QEM definition}
Solution trajectories in the phase-space must obey the set of ODE equations (\ref{d}). Moreover, all trajectories also satisfy a subset of linear equations (\ref{cons.law}) which represent the atom conservation. Among all the concentration points, that fulfill the latter constraints, we can choose those points which minimize the Lyapunov function $G$ of the system we are dealing with. Such points lie on a manifold that is called \textit{Quasi Equilibrium Manifold} (QEM). Let us suppose that some steps of a complex reaction are faster than some others. Since the Lyapunov function G must decrease during the fast dynamics then, when the fast motion is exhausted, the $G$ value is expected to be the minimum on that fast hyperplane. In such a situation, a QEM attempts to achieve a motion decomposition into fast - toward the QEM - and slow - along the QEM. If the invariant manifold exists, the QEM can be taken as a reasonable approximation of it. In order to be more specific, let a chemical system have $n$ reactive species. The degrees of freedom of that system are $(n-l)$ because of the atom balances (\ref{cons.law}). If $q<(n-l)$ is the dimension of the QEM, then the macroscopic variables for its description are $\xi_1,...,\xi_{q}$ so that: $(\mbox{\boldmath$m$}_1,\mbox{\boldmath$c$})=\xi_1,...,(\mbox{\boldmath$m$}_{q},\mbox{\boldmath$c$})=\xi_{q}$. Here, the $n$-dimensional vectors $\mbox{\boldmath$m$}_i$ are related to the hypothetic fast directions. From a mathematical standpoint, the solution of a variational problem:
\begin{equation}\label{QEM_problem}
\left\{ \begin{array}{l}
 G \to min \\ 
 (\mbox{\boldmath$m$}_i ,\mbox{\boldmath$c$}) = \xi _i ,\quad \forall i = 1,..., q \\ 
 \mbox{\boldmath$D$}\mbox{\boldmath$c$} = \mbox{\boldmath$const$} \\ 
 \end{array} \right.
\end{equation}
\begin{figure}
	\centering
		\includegraphics[width=0.70\textwidth]{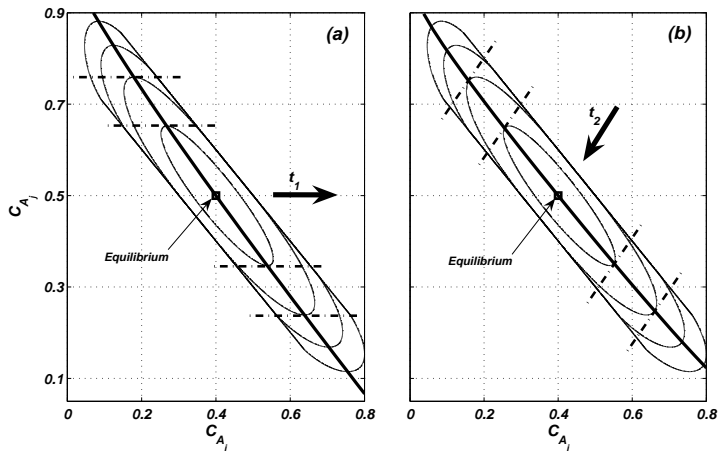}
	\caption{Quasi Equilibrium Manifold: the geometrical interpretation. Two different QE-manifolds (bold lines in (a) and (b)) corresponding to two different linear constraints subsets in the problem (\ref{QEM_problem}).}\label{geom_mean}
\end{figure}
represents the QEM corresponding to the vector set $\{\mbox{\boldmath$m$}_i\}$. We want to stress the geometry behind (\ref{QEM_problem}) because it will be extensively exploited in the following. The geometric interpretation of a QEM is illustrated in Fig. \ref{geom_mean} for a $2$-dimensional phase-space $(c_{A_i},c_{A_j})$. Let us consider the points where $G$ level curves (convex curves in Figures \ref{geom_mean} (a)-(b)) are cut by the QE-manifolds (bold curves): in those points the inclination of the tangent to the $G$-level curves is constant. Different QEM can be obtained by choosing different vector sets $\{\mbox{\boldmath$m$}_i\}$. A special choice is done when $\mbox{\boldmath$m$}_1,...,\mbox{\boldmath$m$}_{q}$ are the $q$ left eigenvectors of the Jacobi matrix $\mbox{\boldmath$L$}(\mbox{\boldmath$c$}^{eq})$ corresponding to the $q$ smallest absolute eigenvalues. In that particular case, solution of (\ref{QEM_problem}) has its own name: \textit{Spectral Quasi Equilibrium Manifold} (SQEM) \cite{1,ChGoKa07}. 

\subsection{Quasi Equilibrium Manifold in practice}
The minimization problem (\ref{QEM_problem}) can be, in principle, solved by the method of Lagrange multipliers. However, it is also well known that, when the number of constraints and variables increases, then that method becomes prohibitively complicated to implement. Since the number of species and elementary reaction steps is usually quite high, the Lagrange multipliers method is not suitable for most of the practical cases. For this reason, in the sequel a new procedure to overcome that issue is presented. An algorithm (\textit{Quasi Equilibrium Grid Algorithm}-QEGA), which can be easily implemented in order to get a discrete analog of a QEM in any dimension, is developed. This is achieved by investigating further the QEM geometrical construction.    

\section{1D Quasi Equilibrium Grid (QEG) construction}\label{main_alg}
\begin{figure}
	\centering
		\includegraphics[width=0.70\textwidth]{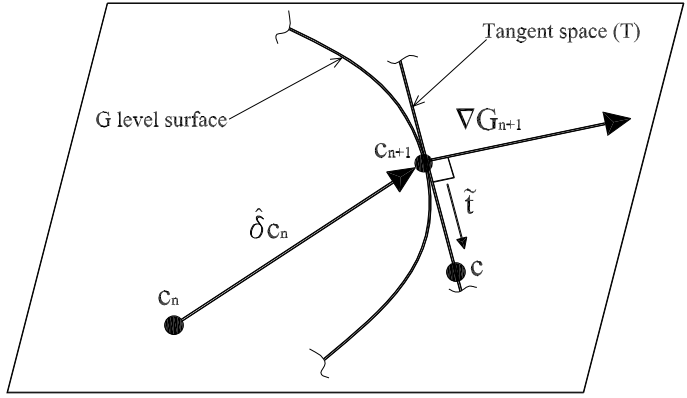}
	\caption{Quasi Equilibrium Grid: the basic idea.}\label{SQEG01}
\end{figure}
Let us consider a one-dimensional quasi-equilibrium manifold. Let us assume that the node $\mbox{\boldmath$c$}_0$ belongs to that manifold. One may now imagine to look for a new node $\mbox{\boldmath$c$}_1$ which still lies on the quasi equilibrium manifold. In general, the node $\mbox{\boldmath$c$}_1$ can be obtained from $\mbox{\boldmath$c$}_0$ by adding a shift $\hat \delta \mbox{\boldmath$c$}_0$: $\mbox{\boldmath$c$}_1  = \mbox{\boldmath$c$}_0  + \hat \delta \mbox{\boldmath$c$}_0$. That idea is applicable whenever a QEM-node $\mbox{\boldmath$c$}_n$ is known and a new one $\mbox{\boldmath$c$}_{n+1}$ must be found (see Fig. \ref{SQEG01}):
\begin{equation}\label{shift}
\mbox{\boldmath$c$}_{n + 1}  = \mbox{\boldmath$c$}_n  + \hat \delta \mbox{\boldmath$c$}_n. 
\end{equation} 
First of all, any node $\mbox{\boldmath$c$}$ has to fulfill the atom balances (\ref{cons.law}). Let $\{\mbox{\boldmath$\rho$}_i\}$ be a basis in the null space of matrix $\mbox{\boldmath$D$}$. A convenient way to take automatically into account the conditions (\ref{cons.law}) is to express any shift $\hat \delta \mbox{\boldmath$c$}_n$ as a linear combination of vectors $\mbox{\boldmath$\rho$}_i$:
\begin{equation}\label{basis_ker}
\hat \delta \mbox{\boldmath$c$}_n  = \sum\nolimits_{i = 1}^z {\mu _i \mbox{\boldmath$\rho$}_i },
\end{equation}
where $z=n-l$ is the dimension of the basis $\{\mbox{\boldmath$\rho$}_i\}$. By referring to Fig. \ref{SQEG01}, let us now discuss further the tangent space $T$ to the $G$ level surface in any quasi-equilibrium point $\mbox{\boldmath$c$}_{n+1}$. The space $T$ geometrically represents the linear constraint of the problem (\ref{QEM_problem}). Therefore, any point $\mbox{\boldmath$c$}$ of $T$ satisfies that constraint, but only $\mbox{\boldmath$c$}_{n+1}$ minimizes $G$ function. The line $\rlap{--} l$ passing from $\mbox{\boldmath$c$}_{n+1}$ and $\mbox{\boldmath$c$}$ has the parametric form $\mbox{\boldmath$c$} = \varphi \mbox{\boldmath$\tilde t$} + \mbox{\boldmath$c$}_{n+1}$, where $\mbox{\boldmath$\tilde t$}$ is a vector of $T$ spanning $\rlap{--} l$ while $\varphi$ is a parameter. In general, the linear constraints of the problem (\ref{QEM_problem}) can be also written as:
\begin{equation}\label{null_space_T}
\left\{ \begin{array}{l}
 \begin{array}{*{20}c}
   {(\mbox{\boldmath$m$},\mbox{\boldmath$c$}) = \varphi (\mbox{\boldmath$m$},\mbox{\boldmath$\tilde t$}) + (\mbox{\boldmath$m$},\mbox{\boldmath$c$}_{n+1})} &  \Rightarrow  & {(\mbox{\boldmath$m$},\mbox{\boldmath$\tilde t$}) = 0,} & {\forall \mbox{\boldmath$\tilde t$}}  \\
\end{array} \\ 
 \begin{array}{*{20}c}
   {(\mbox{\boldmath$d$}_i ,\mbox{\boldmath$c$}) = \varphi (\mbox{\boldmath$d$}_i ,\mbox{\boldmath$\tilde t$}) + (\mbox{\boldmath$d$}_i ,\mbox{\boldmath$c$}_{n+1})} &  \Rightarrow  & {(\mbox{\boldmath$d$}_i ,\mbox{\boldmath$\tilde t$}) = 0,} & {\forall \mbox{\boldmath$\tilde t$}}  \\
\end{array} \\ 
 \end{array} \right.
\end{equation}
where $\mbox{\boldmath$m$}$ and $\mbox{\boldmath$d$}_i$ are the reduced variable vector ($q=1$) and the generic row of matrix $\mbox{\boldmath$D$}$, respectively. The vector $\mbox{\boldmath$\tilde t$}$, that respects (\ref{null_space_T}), can always be written as a linear combination of some vectors $\mbox{\boldmath$t$}_j$, where $\{\mbox{\boldmath$t$}_j\}$ denotes a basis in the null space of that matrix $\mbox{\boldmath$E$}$, whose first row is given by $\mbox{\boldmath$m$}$ and the rest by the rows of $\mbox{\boldmath$D$}$:
\begin{equation}
\mbox{\boldmath$E$} = \left[ \begin{array}{l}
 \mbox{\boldmath$m$} \\ 
 \mbox{\boldmath$D$} \\ 
 \end{array} \right].
\end{equation}
Note that the dimension of $\{\mbox{\boldmath$t$}_j\}$ is $z-1$. By looking at Fig. \ref{SQEG01}, the \textit{quasi-equilibrium requirement} simply becomes the orthogonality condition: 
\begin{equation}\label{orthogonality01}
\begin{array}{*{20}c}
   {(\mbox{\boldmath$\nabla$} G(\mbox{\boldmath$c$}_{n + 1} ),\mbox{\boldmath$\tilde t$}) = 0,} & {\forall \mbox{\boldmath$\tilde t$} \in T}  \\
\end{array}
\end{equation}
which also means:
\begin{equation}\label{orthogonality02}
\begin{array}{*{20}c}
   {(\mbox{\boldmath$\nabla$} G(\mbox{\boldmath$c$}_{n + 1} ),\mbox{\boldmath$t$}_j ) = 0,} & {\forall j = 1,...,z - 1}.  \\
\end{array}
\end{equation}
The quasi-equilibrium grid algorithm is based on the equation system (\ref{orthogonality02}) and two more assumptions. First of all, we suppose that the known node $\mbox{\boldmath$c$}_n$ is close to the QEM, although it does not necessarily belong to the QEM. Secondly, let the vector $\hat \delta \mbox{\boldmath$c$}_n$ be small enough, so that the gradient $\mbox{\boldmath$\nabla$} G(\mbox{\boldmath$c$}_{n+1})$ can be approximated to the first order:
\begin{equation}\label{approx}
\mbox{\boldmath$\nabla$} G(\mbox{\boldmath$c$}_{n + 1})  \cong \mbox{\boldmath$\nabla$} G(\mbox{\boldmath$c$}_n) + \mbox{\boldmath$H$}(\mbox{\boldmath$c$}_n) \hat \delta \mbox{\boldmath$c$}_n, 
\end{equation}
where $\mbox{\boldmath$H$}(\mbox{\boldmath$c$}_n) = \left[ {\frac{{\partial ^2 G}}{{\partial c_i \partial c_j }}} \right]$ denotes again the matrix of second derivatives of the function $G$ evaluated at the point $\mbox{\boldmath$c$}_n$. By substituting equations (\ref{approx}) and (\ref{basis_ker}) in (\ref{orthogonality02}), we obtain:
\begin{equation}\label{sqeg_alg01}
\begin{array}{*{20}c}
   {\sum\nolimits_{i = 1}^z {(\mbox{\boldmath$t$}_j ,\mbox{\boldmath$H$}(\mbox{\boldmath$c$}_n )\mbox{\boldmath$\rho$}_i )} \mu _i  =  - (\mbox{\boldmath$t$}_j ,\mbox{\boldmath$\nabla$} G(\mbox{\boldmath$c$}_n )),} & {\forall j = 1,...,z - 1}  \\
\end{array}.
\end{equation}
By using the entropic scalar product (\ref{entr.prod}), equations (\ref{sqeg_alg01}) can be cast into the form:
\begin{equation}\label{sqeg_alg02}
\begin{array}{*{20}c}
   {\sum\nolimits_{i = 1}^z {\left\langle {\mbox{\boldmath$t$}_j ,\mbox{\boldmath$\rho$}_i } \right\rangle } \mu _i  =  - (\mbox{\boldmath$t$}_j ,\mbox{\boldmath$\nabla$} G(\mbox{\boldmath$c$}_n )),} & {\forall j = 1,...,z - 1}  \\
\end{array}.
\end{equation}
Both the matrix $\mbox{\boldmath$H$}$ and the gradient $\mbox{\boldmath$\nabla$} G$ are calculated at the known node $\mbox{\boldmath$c$}_n$. Note that the right-hand side of (\ref{sqeg_alg02}) vanishes if the node $\mbox{\boldmath$c$}_{n}$ belongs to the QEM. The node collection, subsequently evaluated through (\ref{sqeg_alg02}), will be called a \textit{Quasi Equilibrium Grid} (QEG).

\subsection{Closure through the spacing condition}
\begin{figure}[ht]
	\centering
		\includegraphics[width=0.70\textwidth]{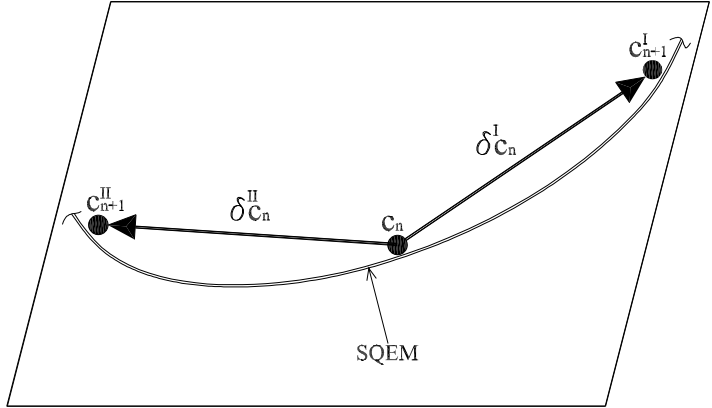}
	\caption{Two solutions for the 1D QEG algorithm.}\label{sol_double}
\end{figure}

Note, however, that the system (\ref{sqeg_alg02}) is not closed ($z$ unknowns $\mu_i$, but $z-1$ equations) because it lacks a further information about the grid spacing. A reasonable closure for that system can be achieved by fixing the grid spacing (e.g. in the Euclidean sense):
\begin{equation}\label{full_qeg_sys}
\left\{ \begin{array}{l}
 \begin{array}{*{20}c}
   {\sum\nolimits_{i = 1}^z {\left\langle {\mbox{\boldmath$t$}_j ,\mbox{\boldmath$\rho$}_i } \right\rangle } \mu _i  =  - (\mbox{\boldmath$t$}_j ,\mbox{\boldmath$\nabla$} G(\mbox{\boldmath$c$}_n )),} & {\forall j = 1,...,z - 1}  \\
\end{array} \\ 
 \left\| {\hat \delta \mbox{\boldmath$c$}_n } \right\| = \varepsilon  \\ 
 \end{array} \right.
\end{equation}
where $\varepsilon$ is a given number and $\left\| {\hat \delta \mbox{\boldmath$c$}_n } \right\|$ represents the Euclidean norm of the vector $\hat \delta \mbox{\boldmath$c$}_n$. The smaller $\varepsilon$ is chosen, the more accurate the expression (\ref{approx}) gets. As it will be shown later on, for small $\varepsilon$ the Quasi Equilibrium Grid lies very close to the correspondent Quasi Equilibrium Manifold.
The extra condition makes (\ref{full_qeg_sys}) a non-linear algebraic system. A way to solve it will be now discussed. The idea is to find the general solution of the linear system (\ref{sqeg_alg02}), and then to choose the one which also fulfills the non linear condition in (\ref{full_qeg_sys}). Let the basis $\{\mbox{\boldmath$\rho$}_i\}$ be orthonormal (in the Euclidean sense). That is not crucial, but it proves to be convenient in the following analysis; indeed the non-linear system (\ref{full_qeg_sys}) now is cast as follows:
\begin{equation}\label{full_sys_normalized}
\left\{ \begin{array}{l}
 \begin{array}{*{20}c}
   {\sum\nolimits_{i = 1}^z {\left\langle {\mbox{\boldmath$t$}_j ,\mbox{\boldmath$\rho$}_i } \right\rangle } \mu _i  =  - (\mbox{\boldmath$t$}_j ,\mbox{\boldmath$\nabla$} G(\mbox{\boldmath$c$}_n )),} & {\forall j = 1,...,z - 1}  \\
\end{array} \\ 
 \sum\nolimits_{i = 1}^z {\mu _i^2 }  = \varepsilon^2.  \\ 
 \end{array} \right.
\end{equation}      
The general solution of (\ref{sqeg_alg02}) can always be written as:
\begin{equation}\label{gen_sol}
\left[ {\begin{array}{*{20}c}
   {\mu _1 }  \\
   {\vdots}   \\
   {\mu _z }  \\
\end{array}} \right] = w\left[ {\begin{array}{*{20}c}
   {\nu_1 }  \\
   {\vdots}  \\
   {\nu_z }  \\
\end{array}} \right] + \left[ {\begin{array}{*{20}c}
   {p_1 }  \\
   {\vdots}\\
   {p_z }  \\
\end{array}} \right],
\end{equation}
where $w$ is a free parameter, while $\mbox{\boldmath$\nu$}=[\nu_1,...,\nu_z]^T$ and $\mbox{\boldmath$p$}=[p_1,...,p_z]^T$ are the solution of the homogeneous problem and a special solution of (\ref{sqeg_alg02}), respectively. Without any restriction, we assume $(\mbox{\boldmath$\nu$},\mbox{\boldmath$\nu$})=1$. Once $\mbox{\boldmath$\nu$}$ and $\mbox{\boldmath$p$}$ are known, the non linear equation of (\ref{full_sys_normalized}) can be written, in terms of $w$, as:
\begin{equation}\label{equat_q}
w^2  + 2(\mbox{\boldmath$\nu$},\mbox{\boldmath$p$})w + (\mbox{\boldmath$p$},\mbox{\boldmath$p$}) - \varepsilon^2  = 0.
\end{equation}
If the solvability condition is satisfied,
\begin{equation}\label{solv_cond}
\left( {\mbox{\boldmath$\nu$} ,\mbox{\boldmath$p$}} \right)^2  - (\mbox{\boldmath$p$},\mbox{\boldmath$p$}) + \varepsilon ^2   > 0,
\end{equation}
then the two real valued solutions of (\ref{equat_q}) ($w^{I}$, $w^{II}$), upon substitution into (\ref{gen_sol}), give two possible sets $[\mu_1,...,\mu_z]$. Therefore, by using the (\ref{shift}) and (\ref{basis_ker}), two new nodes $\mbox{\boldmath$c$}_{n+1}^I, \mbox{\boldmath$c$}_{n+1}^{II}$ (both close to the quasi equilibrium curve) can be evaluated from the previous one $\mbox{\boldmath$c$}_{n}$ (see Fig. \ref{sol_double}).
A criterion, able to choose between those two solutions, depends on the phase-space zone where the grid needs to be constructed. This idea will be clarified in the following example. Usually, the equilibrium point is supposed to be a good starting node for the QEG procedure: $\mbox{\boldmath$c$}_{0}=\mbox{\boldmath$c$}^{eq}$. 

{\it Remark.} The QEG-equations (\ref{sqeg_alg02}) can be generalized as follows:
\begin{equation}
\begin{array}{*{20}c}
   {\sum\nolimits_{i = 1}^z {\left\langle {\mbox{\boldmath$t$}_j ,\mbox{\boldmath$\rho$}_i } \right\rangle } \mu _i  =  - \eta (\mbox{\boldmath$t$}_j ,\mbox{\boldmath$\nabla$} G(\mbox{\boldmath$c$}_n )),} & {\forall j = 1,...,z - 1},  \\
\end{array}
\end{equation}
where $\eta$ is a parameter $0 \le \eta \le 1$. When $\eta=1$, (\ref{sqeg_alg02}) is recovered. On the other hand, if the QEG-nodes are close to the QEM, then the non-homogeneous terms can be neglected (they vanish on the QEM). Therefore, a reasonable approximation of the system (\ref{sqeg_alg02}) is given when $\eta=0$. In the latter case, the solvability condition (\ref{solv_cond}) is fulfilled. If $\eta=1$ and (\ref{solv_cond}) does not hold, that parameter can be chosen in such a way that the solvability condition is satisfied. In the example below, solvability condition (\ref{solv_cond}) is always satisfied and we use (\ref{full_qeg_sys}).  

\section{1D SQEG algorithm at work}\label{ex_1D} 
\begin{figure}
	\centering
		\includegraphics[width=0.90\textwidth]{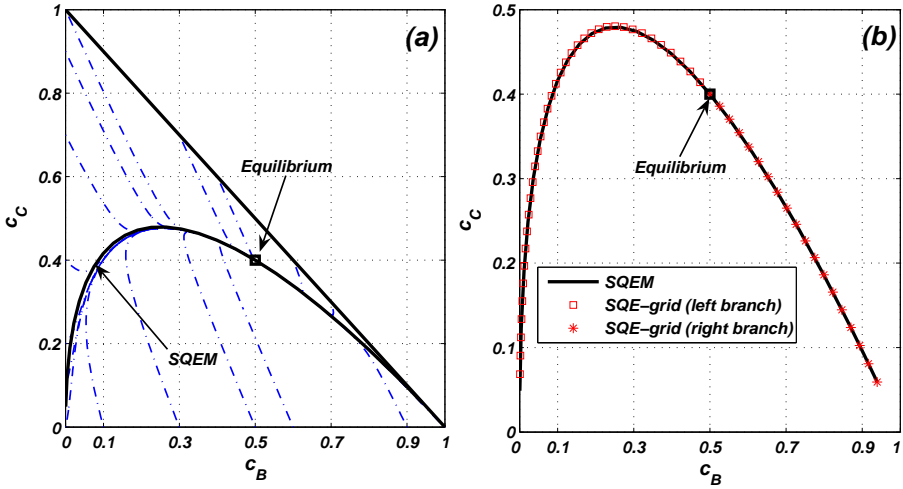}
	\caption{(a) The bold curve is the SQE-manifold which was analytically evaluated by solving the (\ref{SQE_equation}). In that case the SQEM represents a very good approximation of the invariant manifold. (b) The SQE-manifold is compared with the SQE-grid where $\varepsilon^2=10^{-3}$.}\label{grid_manif}
\end{figure}

In this section, an example will be considered in order to illustrate how the algorithm, described in the previous section, works for finding a one-dimensional SQE-grid. That grid will be compared with the relative spectral quasi-equilibrium manifold, too. Let us consider the following four-step three-component reaction (kindly suggested by A.N. Gorban): 
\begin{equation}\label{reaction_1D}
\left\{ \begin{array}{l}
 1.A \leftrightarrow B,\quad k_1^ +   = 1, \\ 
 2.B \leftrightarrow C,\quad k_2^ +   = 1, \\ 
 3.C \leftrightarrow A,\quad k_3^ +   = 1, \\ 
 4.A + B \leftrightarrow 2C,\quad k_4^ +   = 50. \\ 
 \end{array} \right.
\end{equation}
The atom balance takes the form:
\begin{equation}
\mbox{\boldmath$D$}\mbox{\boldmath$c$} = \left[ {\begin{array}{*{20}c}
   1 & 1 & 1  \\
\end{array}} \right]\left[ {\begin{array}{*{20}c}
   {c_A }  \\
   {c_B }  \\
   {c_C }  \\
\end{array}} \right] = 1,
\end{equation}
and the equilibrium point is chosen as: $c_A^{eq}=0.1$, $c_B^{eq}=0.5$, $c_C^{eq}=0.4$.
 \begin{figure}
	\centering
		\includegraphics[width=0.75\textwidth]{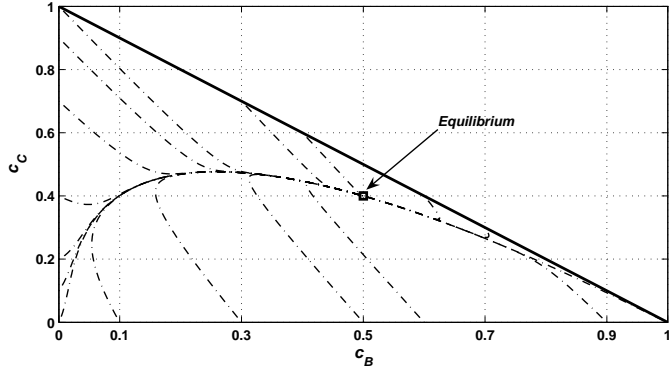}
	\caption{Reaction (\ref{reaction_1D}): some trajectories projected into the phase-subspace $(c_C,c_B)$.}\label{traj_1D}
\end{figure}
As Fig. \ref{traj_1D} shows, the system is effectively two-dimensional, so the quasi-equilibrium manifold is expected to provide an one-dimensional reduced description ($q=1$). Indeed, any solution trajectory, after a rapid initial dynamics, is attracted to a 1D curve and along it reaches the equilibrium point. If the system is closed and the reaction (\ref{reaction_1D}) takes place under constant volume and temperature, we can assume that the system is supported by the Lyapunov function $G$ (\ref{Gfunc}):
\begin{equation}\label{G_func}
G = c_A \left[ {\ln \left( {\frac{{c_A }}{{c_A^{eq} }}} \right) - 1} \right] + c_B \left[ {\ln \left( {\frac{{c_B }}{{c_B^{eq} }}} \right) - 1} \right] + c_C \left[ {\ln \left( {\frac{{c_C }}{{c_C^{eq} }}} \right) - 1} \right].
\end{equation}
Once a 3-dimensional vector $\mbox{\boldmath$m$}$ has been chosen, the QEM equation can be found by solving the variational problem (\ref{QEM_problem}):
\begin{equation}\label{QEM_analytic}
\left\{ \begin{array}{l}
 G \to \min  \\ 
 (\mbox{\boldmath$m$},\mbox{\boldmath$c$}) = \xi \\
 \mbox{\boldmath$D$}\mbox{\boldmath$c$} = 1. \\ 
 \end{array} \right.
\end{equation}
In the following, the Spectral Quasi Equilibrium Manifold (SQEM) \cite{ChGoKa07,1} will be constructed. The Jacobian matrix $\mbox{\boldmath$L$}$ in the equilibrium point and its slowest left eigenvector $\mbox{\boldmath$x$}_l^s$ are:
\begin{equation}\label{Jacob}
\begin{array}{*{20}c}
   {\mbox{\boldmath$L$}(\mbox{\boldmath$c$}^{eq}) = \left[ {\begin{array}{*{20}c}
   { - 30} & { - 4.8} & {13.5}  \\
   { - 24} & { - 6.2} & {13.75}  \\
   {54} & {11} & { - 27.25}  \\
\end{array}} \right],} & {} & {\mbox{\boldmath$x$}_l^s  = \left[ {\begin{array}{*{20}c}
   {0.8807,} & {-0.3905,} & {0.2681}  \\
\end{array}} \right]}.  \\
\end{array}
\end{equation}
Solution of the problem (\ref{QEM_analytic}), with the choice $\mbox{\boldmath$m$}=\mbox{\boldmath$x$}_l^s$, delivers the 1D SQEM for the case shown in Fig \ref{traj_1D}. To this end, let us rewrite the (\ref{QEM_analytic}) in a more explicit form:
\begin{equation}\label{SQEM_sys}
\left\{ \begin{array}{l}
 c_{0A}  = 0.3072 + 0.7867\xi  - 0.5180\phi (\xi ) \\ 
 c_{0B}  = 0.6928 - 0.7867\xi  - 0.4820\phi (\xi ) \\ 
 c_{0C}  = \phi (\xi ) \\ 
 \begin{array}{*{20}c}
   {\frac{{\partial G(\phi ,\xi )}}{{\partial \phi }} = 0,} & {} & {\frac{{\partial ^2 G(\phi ,\xi )}}{{\partial \phi ^2 }} > 0,}  \\
\end{array} \\ 
 \end{array} \right.
\end{equation}
where $\mbox{\boldmath$c$}_0=[c_{0A},c_{0B},c_{0C}]$ is the solution of the problem (\ref{QEM_analytic}), while $\phi$ denotes the relation between $c_C$ and the reduced variable $\xi$ on the SQEM. By using the $G$ function (\ref{G_func}), the problem (\ref{SQEM_sys}) is equivalent to the implicit equation
\begin{equation}\label{SQE_equation}
\left( {\frac{{0.3072 + 0.7867\xi  - 0.5180\phi }}{{0.1}}} \right)^{ - 0.5180} \left( {\frac{{0.6928 - 0.7867\xi  - 0.4820\phi }}{{0.5}}} \right)^{ - 0.4820} \left( {\frac{\phi }{{0.4}}} \right) - 1 = 0.
\end{equation}
The solution of (\ref{SQE_equation}), by means of relations (\ref{SQEM_sys}), gives the SQEM shown in Fig. \ref{grid_manif}(a). One may now apply the QEG-algorithm described above, in order to make a comparison with the analytic solution just found.  
An orthonormal basis $\{\mbox{\boldmath$\rho$}_i\}$ in the null space of the matrix $\mbox{\boldmath$D$}=\left[ {\begin{array}{*{20}c} 1 & 1 & 1  \\ \end{array}} \right]$ has dimension $z=2$ and can be chosen as follows:
\begin{equation}\label{b_vectors}
\left\{ \begin{array}{l}
 \mbox{\boldmath$\rho$}_1  = [-0.5774,0.7887,-0.2113], \\ 
 \mbox{\boldmath$\rho$}_2  = [-0.5774,-0.2113,0.7887]. \\ 
 \end{array} \right.
\end{equation}
Since the matrix $\mbox{\boldmath$E$}$ has the form:
\begin{equation}
\mbox{\boldmath$E$} = \left[ {\begin{array}{*{20}c}
   {0.8807} & {-0.3905} & {0.2680}  \\
   1 & 1 & 1  \\
\end{array}} \right],
\end{equation}
a vector $\mbox{\boldmath$t$}$ spanning ker$(\mbox{\boldmath$E$}$) is:
\begin{equation}\nonumber
	\mbox{\boldmath$t$}=[-0.4229,-0.3934,0.8163].
\end{equation}
The system (\ref{full_sys_normalized}), in this example, simply reads:
\begin{equation}\label{1d_sys}
\left\{ \begin{array}{l}
 \left\langle {\mbox{\boldmath$t$},\mbox{\boldmath$\rho$}_1 } \right\rangle \mu _1  + \left\langle {\mbox{\boldmath$t$},\mbox{\boldmath$\rho$}_2 } \right\rangle \mu _2  =  - \left( {\mbox{\boldmath$t$},\mbox{\boldmath$\nabla$} G} \right) \\ 
 \mu _1^2  + \mu _2^2  = \varepsilon ^2.  \\
 \end{array} \right.
\end{equation}
By solving (\ref{1d_sys}) in a QEG-node $\mbox{\boldmath$c$}_{n}$, the shift vector $\hat \delta \mbox{\boldmath$c$}_{n} = \mu_1 \mbox{\boldmath$\rho$}_1 + \mu_2 \mbox{\boldmath$\rho$}_2$ allows to evaluate the new QEG-node $\mbox{\boldmath$c$}_{n+1}=\mbox{\boldmath$c$}_{n}+\hat \delta \mbox{\boldmath$c$}_{n}$. The QEG procedure, starting from the equilibrium point $\mbox{\boldmath$c$}^{eq}=\mbox{\boldmath$c$}_{0}$, was performed twice, keeping uniformly parameter $\varepsilon^2=10^{-3}$. The first time, by choosing the solution in such a way that $c_{B_{n+1}}<c_{B_n}$, the left branch of the SQE-grid was obtained; then, by imposing $c_{B_{n+1}}>c_{B_n}$, also the right branch was calculated. The algorithm was terminated as soon as at least one component of the new node $\mbox{\boldmath$c$}_{n+1}$ becomes negative.
The result, shown in Fig. \ref{grid_manif}(b), proves that the SQE-grid is in excellent agreement with the analytical curve (SQEM).

\subsection{Grid spacing choice}
\begin{figure}
	\centering
		\includegraphics[width=0.65\textwidth]{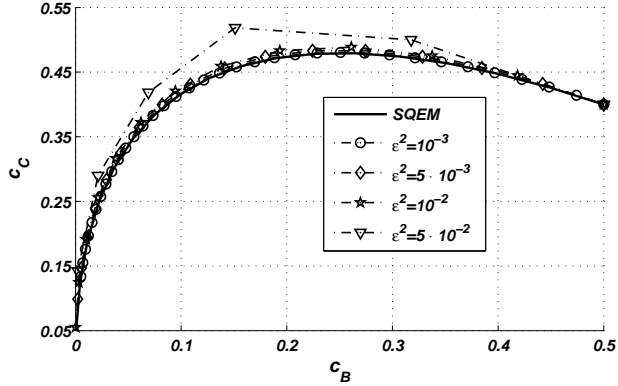}
	\caption{SQEG Left branch of the case in Fig. \ref{grid_manif} (b). Different approximations compared with the analytical solution (SQEM). Each grid is calculated by using a different parameter $\varepsilon$.}\label{accuracy}
\end{figure}
There is no need to stress the importance of the grid spacing parameter $\varepsilon$ for the QEG accuracy. In the case of Section \ref{ex_1D}, the SQEG was computed several times with different values of $\varepsilon$. The QEG algorithm is based on the linear approximation (\ref{approx}). Therefore, the smaller is $\left\| {\hat \delta \mbox{\boldmath$c$}_n} \right\| = \varepsilon$ the more accurate is the QEM description by means of the QEG. Nevertheless, the smaller is $\varepsilon$ the larger is the number of times that the system (\ref{full_qeg_sys}) must be solved to have a grid of a fixed size. For this reason, we need to keep $\varepsilon$ as large as possible. We estimated (at least the order of magnitude) the upper limit of spacing ($\varepsilon_u$) which gives a QEG ``not far'' from the relative QEM. From our numerical experiments, a reasonable value for that was $\varepsilon_u \cong 10^{-1}$. As Fig. \ref{accuracy} shows, the QEG is not far from the QEM even for a quite coarse grid ($\varepsilon>\varepsilon_u$).

\section{Generalization to multi-dimensional grids} \label{Ext_multi}
The QEG algorithm, which has been developed for constructing $1$-dimensional grids, can be modified in order to get multi-dimensional grids, whenever needed. From all reasonable extension strategies, two of them here will be analyzed: a \textit{straightforward extension} and a \textit{flag extension} (flag extension, for invariant grids, was introduced in Ref. \cite{2}). In the first case, the algorithm of paragraph \ref{main_alg} and the equation system (\ref{full_sys_normalized}) are tuned for a $q$-dimensional grid calculation. Here, the implicit assumption is that the grid dimension $q$ is fixed and uniform everywhere in the phase space (like for the QEM construction). However, a second flexible approach, suitable for SQEG construction, was developed, too. In that case, the grid dimension can be varied at will.

\subsection{The straightforward extension}\label{stra_exte}
According to the straightforward extension, if a node $\mbox{\boldmath$c$}_n$ close to the $q$-dimensional QEM is known, then a new node $\mbox{\boldmath$c$}_{n+1}$ can be added to the QE-grid by shifting $\mbox{\boldmath$c$}_n$:
\begin{equation}
\mbox{\boldmath$c$}_{n + 1}  = \mbox{\boldmath$c$}_n  + \hat \delta \mbox{\boldmath$c$}_n ,\quad \hat \delta \mbox{\boldmath$c$}_n  = \sum\nolimits_{i = 1}^z {\mu _i \mbox{\boldmath$\rho$}_i } ,
\end{equation}
where $\{\mbox{\boldmath$\rho$}_i\}$ is still a basis in the null space of matrix $\mbox{\boldmath$D$}$. The linear constraints of the problem (\ref{QEM_problem}) define the tangent space $T$ to the $G$ level surfaces in the new node $\mbox{\boldmath$c$}_{n+1}$. Let $\mbox{\boldmath$c$}$ be a generic point of $T$, the line $\rlap{--} l$ passing from $\mbox{\boldmath$c$}_{n+1}$ and $\mbox{\boldmath$c$}$ has the parametric form: $\mbox{\boldmath$c$}=\varphi \mbox{\boldmath$\tilde t$}+\mbox{\boldmath$c$}_{n+1}$, where $\mbox{\boldmath$\tilde t$}$ is the vector of $T$ which spans $\rlap{--} l$ and $\varphi$ is the parameter. The generalized form of the relations (\ref{null_space_T}) is:
\begin{equation}
\left\{ \begin{array}{l}
 \left( {\mbox{\boldmath$m$}_1 ,\mbox{\boldmath$c$}} \right) = \varphi \left( {\mbox{\boldmath$m$}_1 ,\mbox{\boldmath$\tilde t$}} \right) + \left( {\mbox{\boldmath$m$}_1 ,\mbox{\boldmath$c$}_{n + 1}^i } \right) \Rightarrow \left( {\mbox{\boldmath$m$}_1 ,\mbox{\boldmath$\tilde t$}} \right) = 0,\quad \forall \mbox{\boldmath$\tilde t$} \in T \\ 
  \vdots  \\ 
 \left( {\mbox{\boldmath$m$}_q ,\mbox{\boldmath$c$}} \right) = \varphi \left( {\mbox{\boldmath$m$}_q ,\mbox{\boldmath$\tilde t$}} \right) + \left( {\mbox{\boldmath$m$}_q ,\mbox{\boldmath$c$}_{n + 1}^i } \right) \Rightarrow \left( {\mbox{\boldmath$m$}_q ,\mbox{\boldmath$\tilde t$}} \right) = 0,\quad \forall \mbox{\boldmath$\tilde t$} \in T \\ 
 \left( {\mbox{\boldmath$d$}_i ,\mbox{\boldmath$c$}} \right) = \varphi \left( {\mbox{\boldmath$d$}_i ,\mbox{\boldmath$\tilde t$}} \right) + \left( {\mbox{\boldmath$d$}_i ,\mbox{\boldmath$c$}_{n + 1}^i } \right) \Rightarrow \left( {\mbox{\boldmath$d$}_i ,\mbox{\boldmath$\tilde t$}} \right) = 0,\quad \forall \mbox{\boldmath$\tilde t$} \in T, \\ 
 \end{array} \right.
\end{equation}
which means that vector $\mbox{\boldmath$\tilde t$}$ belongs to the null space of the matrix $\mbox{\boldmath$E$}$ (ker$\mbox{\boldmath$E$}$):
\begin{equation}
\mbox{\boldmath$E$} = \left[ \begin{array}{l}
 \mbox{\boldmath$m$}_1  \\ 
  \vdots  \\ 
 \mbox{\boldmath$m$}_q  \\ 
 \mbox{\boldmath$D$} \\
 \end{array} \right].
\end{equation}
Now, the dimension of basis $\{\mbox{\boldmath$t$}_j\}$ in ker$(\mbox{\boldmath$E$})$ is ($z-q$). Since the quasi equilibrium condition requires that, among all the points $\mbox{\boldmath$c$}$ of $T$, $\mbox{\boldmath$c$}_{n+1}$ has the minimal value of $G$, the following orthogonality conditions hold:
\begin{equation}\label{gen_orthog}
 \left( {\mbox{\boldmath$\nabla$} G(\mbox{\boldmath$c$}_{n + 1}),\mbox{\boldmath$t$}_j } \right) = 0,\quad \forall j = 1,...,z - q.
\end{equation}
For small vector $\hat \delta \mbox{\boldmath$c$}_n$, the approximation (\ref{approx}) can be used, so that the (\ref{gen_orthog}) become:
\begin{equation}\label{QEG_general}
 \sum\nolimits_{i = 1}^z {\left\langle {\mbox{\boldmath$t$}_j ,\mbox{\boldmath$\rho$}_i } \right\rangle \mu _i  =  - \left( {\mbox{\boldmath$t$}_j ,\mbox{\boldmath$\nabla$} G(\mbox{\boldmath$c$}_n )} \right),\quad \forall j = 1,...,z - q}.  \\ 
\end{equation}
As the system (\ref{QEG_general}) shows, the larger is the QEM dimension ($q$) the smaller is the set of ``mere'' quasi-equilibrium equations available, while the number of unknowns remains constant ($z$). The closure of the rectangular system (\ref{QEG_general}) requires $q$ more equations and has only to do with the geometric structure which we want to provide the grid with (e.g. grid spacing, shift vector orientation in the phase-space, etc). In general, the geometric structure of the grid under construction can be chosen at will: therefore there is no unique geometric closure for that system. However, one possible condition could be imposed, like in (\ref{full_qeg_sys}), by fixing the Euclidean norm of shift vector: $\left\| {\hat \delta \mbox{\boldmath$c$}_n } \right\| = \varepsilon$. Nevertheless, ($q-1$) geometric constraints are still missing. In order to illustrate how the geometric closure issue can be overcome, the case $q=2$ will be considered in the following. For that special case, a possible closure, which can be easily generalized, will be presented. If a two-dimensional QEG has to be constructed, then only one extra equation is needed to close the system:
\begin{equation}\label{system2D}
\left\{ \begin{array}{l}
 \sum\nolimits_{i = 1}^z {\left\langle {\mbox{\boldmath$t$}_j ,\mbox{\boldmath$\rho$}_i } \right\rangle \mu _i  =  - \left( {\mbox{\boldmath$t$}_j ,\mbox{\boldmath$\nabla$} G(\mbox{\boldmath$c$}_n )} \right),\quad \forall j = 1,...,z - 2}  \\ 
 \left\| {\hat \delta \mbox{\boldmath$c$}_n } \right\| = \varepsilon.  \\ 
 \end{array} \right.
\end{equation}
\begin{figure}
	\centering
		\includegraphics[width=0.70\textwidth]{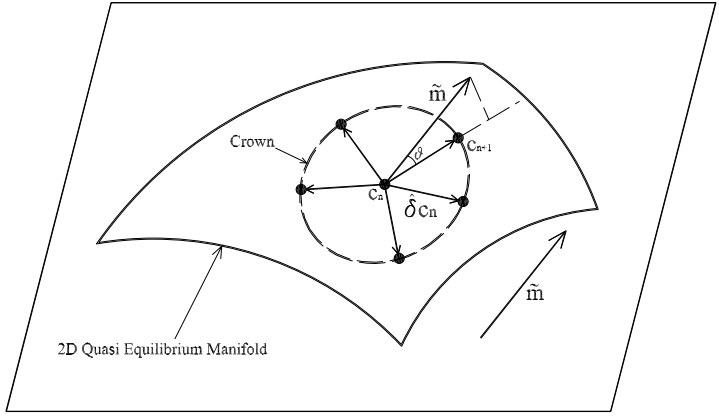}
	\caption{2D Quasi Equilibrium Manifold. Location of solutions of the system (\ref{system2D}) in the phase space.}\label{SQEG2D}
\end{figure}
Fig. \ref{SQEG2D} shows that all the possible solutions of (\ref{system2D}) are located, as a ``crown'', near the QEM. A way to choose only two of them can be achieved by introducing a new fixed vector $\mbox{\boldmath$\tilde m$}$ and imposing a given angle $\vartheta$ between $\mbox{\boldmath$\tilde m$}$ and $\hat \delta \mbox{\boldmath$c$}_n$:
\begin{equation}\label{extra_cond}
\sum\nolimits_{i = 1}^z { \left(\mbox{\boldmath$\tilde m$},{\mbox{\boldmath$\rho$}_i} \right)\mu _i}  = \left\| {\hat \delta \mbox{\boldmath$c$}_n } \right\| \cdot \left\| {\mbox{\boldmath$\tilde m$}} \right\|\cos \vartheta. 
\end{equation}
The choice $\vartheta=\pi/2$ proves to be particularly convenient, as (\ref{extra_cond}) becomes:
\begin{equation}\label{short_extra_cond}
\sum\nolimits_{i = 1}^z {\left( {\mbox{\boldmath$\tilde m$},\mbox{\boldmath$\rho$}_i } \right)\mu _i }  = 0.	
\end{equation}
(\ref{short_extra_cond}) allows to write a closed system:
\begin{equation}\label{closed_system_2D}
\left\{ \begin{array}{l}
 \sum\nolimits_{i = 1}^z {\left\langle {\mbox{\boldmath$t$}_j ,\mbox{\boldmath$\rho$}_i } \right\rangle \mu _i  =  - \left( {\mbox{\boldmath$t$}_j ,\mbox{\boldmath$\nabla$} G(\mbox{\boldmath$c$}_n )} \right),\quad \forall j = 1,...,z - 2}  \\ 
 \sum\nolimits_{i = 1}^z { \left( {\mbox{\boldmath$\tilde m$} , \mbox{\boldmath$\rho$}_i } \right)\mu _i}  = 0 \\ 
 \left\| {\hat \delta \mbox{\boldmath$c$}_n } \right\| = \varepsilon,  \\ 
 \end{array} \right.  
\end{equation}
where the extra information, through $\varepsilon$ and $\mbox{\boldmath$\tilde m$}$, concerns the grid spacing and the phase-space zone of interest where the grid must be constructed. In general, the geometric closure of (\ref{QEG_general}) can be achieved when $(q-1)$ independent vectors $\{\mbox{\boldmath$\tilde m$}_i\}$ and the parameter $\varepsilon$ are fixed. Here, we present an approach which allows to get a rectangular structured grid. The general form of (\ref{closed_system_2D}) is:
\begin{equation}\label{closed_system_qD}
\left\{ \begin{array}{l}
 \sum\nolimits_{i = 1}^z {\left\langle {\mbox{\boldmath$t$}_j ,\mbox{\boldmath$\rho$}_i } \right\rangle \mu _i  =  - \left( {\mbox{\boldmath$t$}_j ,\mbox{\boldmath$\nabla$} G(\mbox{\boldmath$c$}_n )} \right),\quad \forall j = 1,...,z - q}  \\ 
 \sum\nolimits_{i = 1}^z { \left( {\mbox{\boldmath$\tilde m$}_j , \mbox{\boldmath$\rho$}_i } \right)\mu _i}  = 0,\quad  \forall j = 1,...,q-1\\ 
 \left\| {\hat \delta \mbox{\boldmath$c$}_n } \right\| = \varepsilon.  \\ 
 \end{array} \right.	
\end{equation}
The $q$-dimensional grid construction is split in $q$ subsequent steps. Starting from the equilibrium $\mbox{\boldmath$c$}^{eq}$, system (\ref{closed_system_qD}) is solved by choosing ($q-1$) $\mbox{\boldmath$m$}_j$ vectors among the $q$ available and imposing: $\mbox{\boldmath$\tilde m$}_j=\mbox{\boldmath$m$}_j \quad \forall j=1,...,q-1$. In this way, a first set of QEG nodes is attained as soon as $\varepsilon$ is known. Now, starting from each of those points, system (\ref{closed_system_qD}), by using a different combination of $\mbox{\boldmath$m$}_j$ vectors, gives some more nodes. The procedure ends ($q$-th step) when all the possible different combinations of ($q-1$) vectors $\{\mbox{\boldmath$m$}_j\}$ are over. In Section \ref{2D_example}, that idea will be explained by means of an illustrative example. 

\subsection{The flag extension}\label{flag_ext}
\begin{figure}
	\centering
		\includegraphics[width=0.70\textwidth]{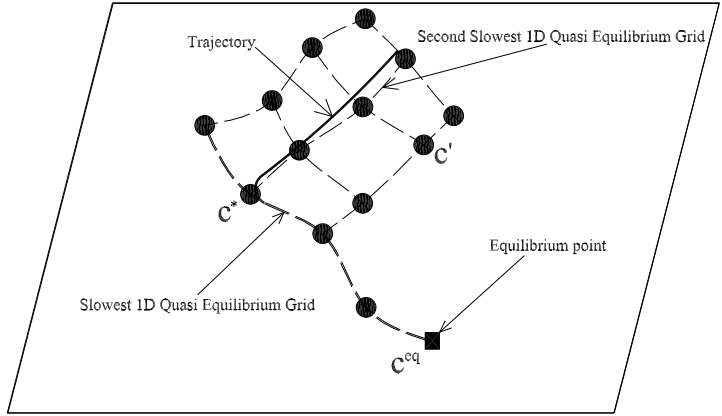}
	\caption{A 2D flag. Once the 1D quasi equilibrium grid is found, from each node $\mbox{\boldmath$c$}^*$, new 1D quasi equilibrium grids are added. The second slowest 1D grid represents that trajectory collected by the first 1D quasi equilibrium grid in the node $\mbox{\boldmath$c$}^*$.}\label{SQEG_2D_flag}
\end{figure}

A multi-dimensional QE-grid construction becomes non-trivial especially when $q$ becomes large. As reported in section \ref{stra_exte}, the straightforward extension requires to introduce some additional vectors $\mbox{\boldmath$\tilde m$}_j$. The flag extension can be applied when a SQEG is searched. That procedure is strongly based on the algorithm presented in paragraph \ref{main_alg} and it naturally leads to a rectangular structured grid. The idea, which is behind, is simple and makes this method very flexible and really suitable for constructing high-dimensional rectangular grids. Let us suppose that $q$ is the grid dimension and the $q$ SQE-vectors $\{\mbox{\boldmath$m$}_1,...,\mbox{\boldmath$m$}_s\}$ are fixed. Here, the assumption is that $\mbox{\boldmath$m$}_1$ is the slowest eigenvector (corresponding to the smallest eigenvalue by absolute value), $\mbox{\boldmath$m$}_2$ the second slowest and so forth. The grid construction is achieved in $s$ subsequent steps. In each step one more dimension is added to the grid. At the beginning, by using $\mbox{\boldmath$m$}=\mbox{\boldmath$m$}_1$, the algorithm in section \ref{main_alg} provides the 1D quasi-equilibrium grid. Now, starting from any node $\mbox{\boldmath$c$}^*$ of that grid, a new 1D QEG is constructed where $\mbox{\boldmath$m$}=\mbox{\boldmath$m$}_2$. In this case, the second QEG represents a trajectory on the 2D-manifold attracted to the slowest 1D-manifold in the node $\mbox{\boldmath$c$}^*$, once the fast dynamics is exhausted (see Fig. \ref{SQEG_2D_flag}). $G$ function depends on the equilibrium point $\mbox{\boldmath$c$}^{eq}$: $G=G(\mbox{\boldmath$c$},\mbox{\boldmath$c$}^{eq})$. Since $\mbox{\boldmath$c$}^*$ can be considered as a ``local equilibrium'' for the fast motion, the second 1D grid is obtained by minimizing $G=G(\mbox{\boldmath$c$},\mbox{\boldmath$c$}^{*})$. Once the previous step is completed, the grid can be extended in the third dimension by adding, in each node $\mbox{\boldmath$c$}^{'}$ of the new 2D grid, a 1D QEG where $\mbox{\boldmath$m$}=\mbox{\boldmath$m$}_3$ and $G=G(\mbox{\boldmath$c$},\mbox{\boldmath$c$}^{'})$. In this way, the procedure is performed up to a $q$-dimensional grid. By extending partially a previous grid, it becomes really easy to have some grids whose dimension is different in different phase space zones. It is worth to stress that the straightforward and the flag extension deliver two different objects: the first one just gives the quasi-equilibrium grid ``brute force'', while the second one is its convenient ``approximation'' which has some useful features as it will be illustrated in the following.
First of all, the flag grid does not demand any extra vector for the geometric closure and the grid dimension can be easily varied in different phase-space zones. Secondly, if a grid refinement procedure (MIG) is used in order to get an invariant grid out of the quasi-equilibrium one \cite{ChGoKa07}, then the flag extension reveals to be a very useful tool. Indeed, let us assume that a multi-dimensional invariant grid is required in order to reduce a given model. A possible strategy might be given by a ``hybrid procedure'' where the QEG algorithm and the MIG method are alternatively used according to the sequence:
\begin{itemize}
	\item 1D quasi-equilibrium grid construction (slowest grid);
	\item MIG refinements until the 1D invariant grid is obtained;
	\item flag extension from 1D invariant grid to 2D quasi-equilibrium grid;
	\item MIG refinements until the 2D invariant grid is obtained;
	\item flag extension from 2D invariant grid to 3D quasi-equilibrium grid;
	\item MIG refinements...
\end{itemize}

\subsection{Beyond SQEG: GQEG and SEGQEG}\label{Beyond}
The latter suggestion sheds light on one more option which, if implemented during the flag extension, allows to go beyond the SQEG approximation of the invariant manifold. Let us assume that the hybrid procedure of Section $\ref{flag_ext}$ is utilized and a $k$-dimensional invariant grid (let $\mbox{\boldmath$c$}^*$ be its generic node) has to be extended to a ($k+1$)-dimensional grid. That grid will approximate the ($k+1$)-dimensional invariant grid, better than the SQEG does, if in each invariant node $\mbox{\boldmath$c$}^*$ the vector $\mbox{\boldmath$m$}$ is chosen as the ($k+1$)-th slowest left eigenvector (by absolute value) of Jacobi matrix $\mbox{\boldmath$L$}(\mbox{\boldmath$c$}^*)$. According to \cite{1}, here a considerable simplification can be achieved by replacing the full Jacobian $\mbox{\boldmath$L$}(\mbox{\boldmath$c$}^*)$ with:
\begin{equation}\label{sym_entr_jac}
	\mbox{\boldmath$L$}^{sym}(\mbox{\boldmath$c$}^*)=\frac{1}{2}\left( {\mbox{\boldmath$L$}(\mbox{\boldmath$c$}^* ) + \mbox{\boldmath$H$}^{ - 1} \mbox{\boldmath$L$}^T (\mbox{\boldmath$c$}^* )\mbox{\boldmath$H$}} \right),
\end{equation}
where $\mbox{\boldmath$L$}^T$ is the ordinary transposition, and $\mbox{\boldmath$H$}$ is evaluated at the point $\mbox{\boldmath$c$}^*$, too. Matrix $\mbox{\boldmath$L$}^{sym}$ is symmetric with respect to the entropic scalar product (\ref{entr.prod}). For that reason the spectral decomposition will be much more viable (see also Ref. \cite{ChGoKa07}). Those two new approximations will be named: \textit{Guided Quasi Equilibrium Grid} (GQEG) when the full Jacobian $\mbox{\boldmath$L$}(\mbox{\boldmath$c$}^*)$ is used, while \textit{Symmetric Entropic Guided Quasi Equilibrium Grid} (SEGQEG) if $\mbox{\boldmath$L$}^{sym}$ replaces the full matrix. In order to give an idea about the effort needed, for example in a SEGQEG construction, let us consider a 2-dimensional grid. In that case, the spectral decomposition of a symmetric operator is performed only over the nodes of an one-dimensional grid. Moreover, also a possible criterion, for getting a multi-dimensional grid, naturally applies: if at the node $\mbox{\boldmath$c$}^*$ of the $k$-dimensional invariant grid, the ratio ${{\left| {\lambda _{k + 1} } \right|} \mathord{\left/
 {\vphantom {{\left| {\lambda _{k + 1} } \right|} {\left| {\lambda _k } \right|}}} \right.
 \kern-\nulldelimiterspace} {\left| {\lambda _k } \right|}}$ (between eigenvalues of $\mbox{\boldmath$L$}$ or $\mbox{\boldmath$L$}^{sym}$, respectively) is not larger than a fixed threshold, the ($k+1$)-dimensional grid will not be extended at that point. In this way, the grid dimension $q$ is generally not uniform in the phase space. Several techniques suggested above are only some reasonable ones. The flexibility of the method proposed allows to set up different procedures, still based on the Quasi Equilibrium Grid approach: the QEG system (\ref{QEG_general}) supplied by a geometrical closure.
   
\section{2D Grid Example: hydrogen oxidation reaction}\label{2D_example}
Let us consider a model for hydrogen oxidation reaction where six species $H_2$ (hydrogen), $O_2$ (oxygen), $H_2O$ (water), $H$, $O$, $OH$ (radicals) are involved in six steps in a closed system under constant volume and temperature (see Ref. \cite{book}, p. 291): 
\begin{equation}\label{hydrogen_model}
\left\{ \begin{array}{l}
 1.H_2  \leftrightarrow 2H,\quad k_1^ + = 2, \\ 
 2.O_2  \leftrightarrow 2O,\quad k_2^ + = 1, \\ 
 3.H_2 O \leftrightarrow H + OH,\quad k_3^ + = 1, \\ 
 4.H_2  + O \leftrightarrow H + OH,\quad k_4^ + = 10^3 , \\ 
 5.O_2  + H \leftrightarrow O + OH,\quad k_5^ + = 10^3 , \\ 
 6.H_2  + O \leftrightarrow H_2 O,\quad k_6^ + = 10^2 . \\ 
 \end{array} \right.
\end{equation}
The conservation laws are:
\begin{equation}
\left\{ \begin{array}{l}
 2c_{H_2 }  + 2c_{H_2 O}  + c_H  + c_{OH}  = b_H  = 2 \\ 
 2c_{O_2 }  + c_{H_2 O}  + c_O  + c_{OH}  = b_O  = 1. \\ 
 \end{array} \right.
\end{equation}
When the equilibrium point is fixed, for example
\begin{equation}\label{eq_coord_2D}
c_{H_2 }^{eq}  = 0.27,\;c_{O_2 }^{eq}  = 0.135,\;c_{H_2 O}^{eq}  = 0.7,\;c_H^{eq}  = 0.05,\;c_O^{eq}  = 0.02,\;c_{OH}^{eq}  = 0.01,
\end{equation}
then the rest of the rate constants $k_i^-$ are calculated using the detailed balance principle (\ref{detail.balance}). The system under consideration is fictitious in the sense that the subset of equations corresponds to the simplified picture of this chemical process and the rate constants reflect only orders of magnitude for relevant real-word systems. We can assume that the Lyapunov function $G$ has the form:
\begin{equation}\label{hydrogen_lyapunov}
G = \sum\nolimits_{i = 1}^6 {c_i \left[ {\ln \left( {\frac{{c_i }}{{c_i^{eq} }}} \right) - 1} \right]}. 
\end{equation}
Here, we are interested in the 2D SQEG construction. Two left eigenvectors of Jacobian matrix $\mbox{\boldmath$L$}(\mbox{\boldmath$c$}^{eq})$ are:
\begin{equation}
\left\{ \begin{array}{l}
 \mbox{\boldmath$x$}_l^{s1}  = \left[ { - 0.577, - 0.568,0.225,0.0482,0.0666, - 0.536} \right] \\ 
 \mbox{\boldmath$x$}_l^{s2}  = \left[ {0.00682,-0.00595,0.0221,-0.7,-0.713,0.423} \right], \\ 
 \end{array} \right.
\end{equation}
where $\mbox{\boldmath$x$}_l^{s1}$ and $\mbox{\boldmath$x$}_l^{s2}$ are the slowest and the second slowest one, respectively. 

\subsection{The 2D straightforward extension}
\begin{figure}
	\centering
		\includegraphics[width=0.80\textwidth]{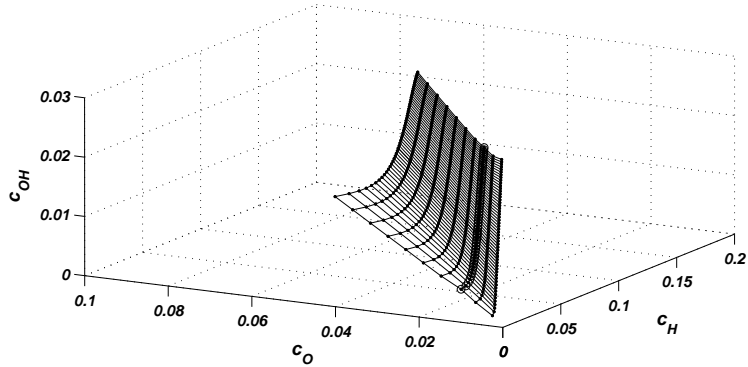}
	\caption{The 2D SQEG constructed by using the straightforward extension with $\varepsilon^2=0.5 \cdot 10^{-3}$: projection into the phase-subspace ($c_H,c_O,c_{OH}$).}\label{SQEG_2D_stra}
\end{figure}
In order to get a 2D SQEG for that example, the straightforward extension was used as first strategy. Matrices $\mbox{\boldmath$D$}$ and $\mbox{\boldmath$E$}$ take now the form:
\begin{equation} 
\mbox{\boldmath$D$} = \left[ {\begin{array}{*{20}c}  
   2 & 0 & 2 & 1 & 0 & 1  \\
   0 & 2 & 1 & 0 & 1 & 1  \\
\end{array}} \right],{\kern 1pt} \mbox{\boldmath$E$} = \left[ {\begin{array}{*{20}c}
   {- 0.577} & {- 0.568} & {0.225} & {0.0482} & { - 0.0666} & { - 0.536}  \\
   {0.00682} & { - 0.00595} & {0.0221} & { - 0.7} & { - 0.713} & {0.423}  \\
   2 & 0 & 2 & 1 & 0 & 1  \\
   0 & 2 & 1 & 0 & 1 & 1  \\
\end{array}} \right].
\end{equation}
As suggested in the end of section \ref{stra_exte}, the procedure has been started from the equilibrium point and it was split in two subsequent steps. At the beginning, the system (\ref{closed_system_2D}) was solved by imposing $\varepsilon^2=0.5 \cdot 10^{-3}$ and $\mbox{\boldmath$\tilde m$}=\mbox{\boldmath$x$}_l^{s2}$: in this way, the grid nodes, denoted by circles, in Fig. \ref{SQEG_2D_stra} were obtained. In the second step, (\ref{closed_system_2D}) was solved by starting from any circle: this time, the geometric constraints were $\varepsilon^2=0.5 \cdot 10^{-3}$ and $\mbox{\boldmath$\tilde m$}=\mbox{\boldmath$x$}_l^{s1}$. During that step, in each circle, the horizontal dots of Fig. \ref{SQEG_2D_stra} were found, too.

\subsection{The 2D flag extension}
\begin{figure}
	\centering
		\includegraphics[width=0.80\textwidth]{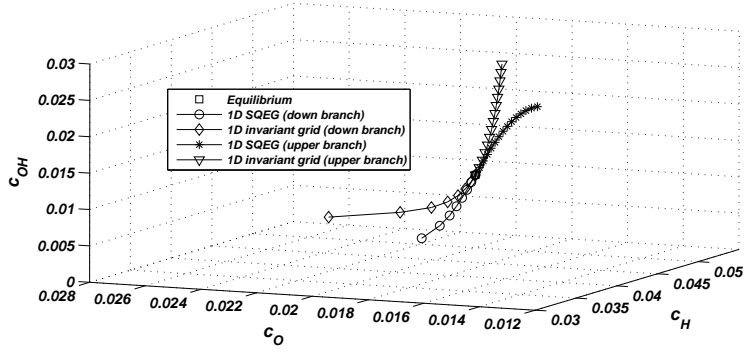}
	\caption{1D Spectral Quasi Equilibrium Grid with $\varepsilon^2=3 \cdot 10^{-3}$: comparison with the 1D invariant grid obtained by MIG refinements.}\label{Hydr_1D_comp}
\end{figure}

After that, also the flag extension procedure was applied. Now, the 1D spectral quasi equilibrium grid is needed. Matrices $\mbox{\boldmath$D$}$ and $\mbox{\boldmath$E$}$ are in this case:
\begin{equation}
\mbox{\boldmath$D$} = \left[ {\begin{array}{*{20}c}
   2 & 0 & 2 & 1 & 0 & 1  \\
   0 & 2 & 1 & 0 & 1 & 1  \\
\end{array}} \right],{\kern 1pt} \mbox{\boldmath$E$} = \left[ {\begin{array}{*{20}c}
   { - 0.577} & { - 0.568} & {0.225} & {0.0482} & {0.0666} & { - 0.536}  \\
   2 & 0 & 2 & 1 & 0 & 1  \\
   0 & 2 & 1 & 0 & 1 & 1  \\
\end{array}} \right].
\end{equation}
Starting from the equilibrium point $\mbox{\boldmath$c$}^{eq}$, the system (\ref{full_sys_normalized}) was solved by fixing $\varepsilon^2=3\cdot10^{-3}$ (see Fig. \ref{Hydr_1D_comp}).
The flag extension was used to get a 2D grid out of the 1D one. Now, the new matrix $\mbox{\boldmath$E$}$ reads:
\begin{figure}
	\centering
		\includegraphics[width=0.80\textwidth]{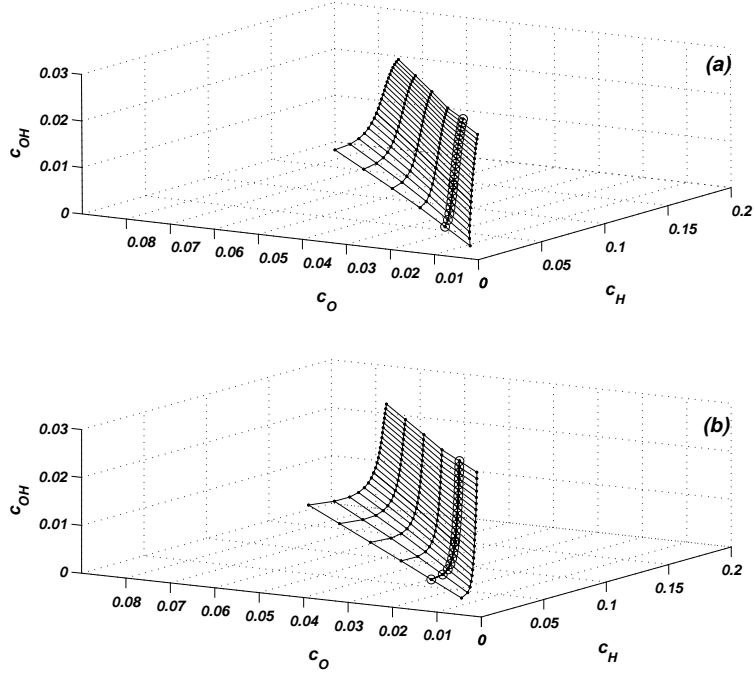}
	\caption{The flag extension. (a) 2D SQEG (dots) extended from the 1D SQEG (circles). (b) 2D SQEG (dots) extended from the 1D invariant grid (circles). The grid spacing, in the second dimension, was $\varepsilon^2=1.5 \cdot 10^{-3}$.}\label{SQEG_2D_flag_01}
\end{figure}
\begin{equation}\nonumber
\mbox{\boldmath$E$} = \left[ {\begin{array}{*{20}c}
   {0.00682} & {-0.00595} & {0.0221} & {-0.7} & {-0.713} & {0.423}  \\
   2 & 0 & 2 & 1 & 0 & 1  \\
   0 & 2 & 1 & 0 & 1 & 1  \\
\end{array}} \right],
\end{equation}
while the Lyapunov function $G$ has the form:
\begin{equation}
G = \sum\nolimits_{i = 1}^6 {c_i \left[ {\ln \left( {\frac{{c_i }}{{c_i^* }}} \right) - 1} \right]}, 
\end{equation}
where $\mbox{\boldmath$c$}^*=[c_1^*,...,c_6^*]$ is any 1D grid node which is extended in the second dimension (see Fig. \ref{SQEG_2D_flag}).
Figures \ref{SQEG_2D_flag_01}(a)-(b) show two different 2D SQE-grids: the first one is obtained by extending the 1D SQE-grid, while in the second case the 1D invariant grid is used. In other words, the latter result was attained by the ``hybrid procedure'' $QEGA$ + $MIG$ suggested in the end of section \ref{flag_ext}. For both cases, in the second dimension, the grid spacing was $\varepsilon^2=1.5 \cdot 10^{-3}$.

\subsection{The 2D GQEG and SEGQEG}\label{compar_2D} 
\begin{figure}[t]
	\centering
		\includegraphics[width=0.80\textwidth]{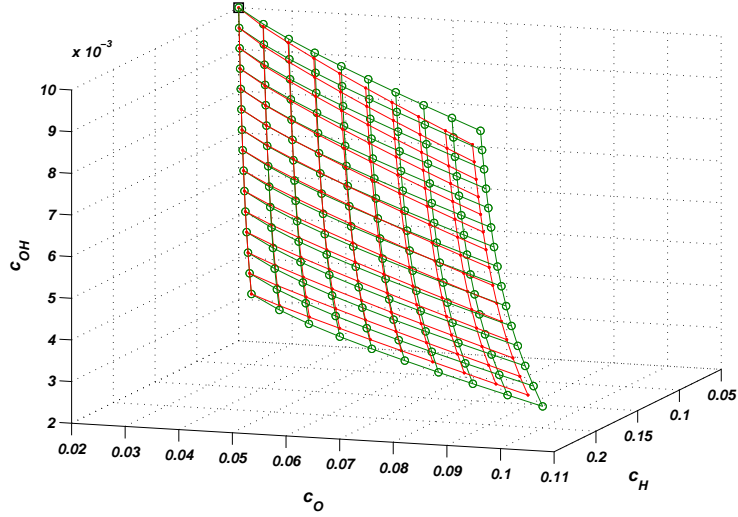}
	\caption{1.case: $k_1^+=2$, $k_2^+=1$, $k_3^+=1$, $k_4^+=10^3$, $k_5^+=10^3$, $k_6^+=10^2$, $\varepsilon ^2  = 0.45 \cdot 10^3$. Two grids formed by $10 \times 15$ nodes. A 2D GQEG (dots) and a 2D invariant grid (circles) are reported. The grids are partially extended below the equilibrium point (square).}\label{comparison_2D_01}
\end{figure}
Finally, the GQEG and SEGQEG approximations are computed for the hydrogen oxidation reaction (\ref{hydrogen_model}) (1.case). Here, the grid spacing is uniformly kept $\varepsilon^2=0.45 \cdot 10^{-3}$. Each grid has $10 \times 15$ nodes and it is compared with both the SQEG (straightforward extension) of similar size (check Table \ref{table01}) and the invariant one. The invariant grid was obtained by refining the approximations through the MIG procedure. All those grids lie quite close to each other. However, a ``more pathological'' case (2.case) is also analyzed (here the SQEG, far from the equilibrium, presents a remarkable deviation from the invariant grid): now the rate constant set is taken as $k_1^+=20$, $k_2^+=1$, $k_3^+=1$, $k_4^+=10^3$, $k_5^+=10^3$, $k_6^+=10^2$, while the equilibrium point coordinates still are given by (\ref{eq_coord_2D}).
\begin{figure}
	\centering
		\includegraphics[width=0.80\textwidth]{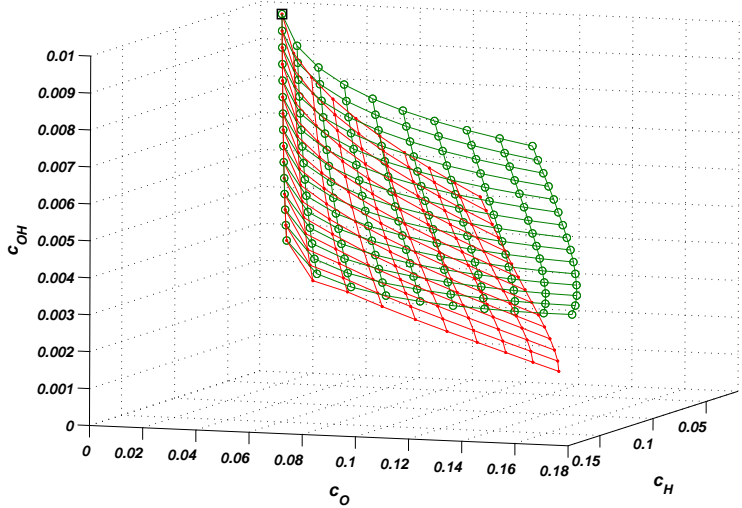}
	\caption{2.case: $k_1^+=20$, $k_2^+=1$, $k_3^+=1$, $k_4^+=10^3$, $k_5^+=10^3$, $k_6^+=10^2$, $\varepsilon ^2  = 0.45 \cdot 10^3$. Two grids formed by $10 \times 15$ nodes. A 2D GQEG (dots) and a 2D invariant grid (circles) are reported. The grids are partially extended below the equilibrium point (square).}\label{comparison_2D_02}
\end{figure}
For that case, the SQEG, GQEG and SEGQEG were constructed by choosing the grid spacing and size as for the previous case. Note that all the grids were partially extended only below the equilibrium point in the phase-space zone where they present the largest deviation from the invariant grid. This time those three approximations have a low invariance defect only near the equilibrium. In order to estimate how far each grid is from the invariant one, the following procedure is implemented. A $10 \times 15$ matrix, collecting in any grid node an invariance defect measure, is constructed. As suggested by \cite{ChGoKa07}, that local measure may be $\sqrt {\left( {\mbox{\boldmath$\Delta$} ,\mbox{\boldmath$\Delta$} } \right)/\left( {\mbox{\boldmath$J$},\mbox{\boldmath$J$}} \right)}$, where $\mbox{\boldmath$\Delta$}$ and $\mbox{\boldmath$J$}$ are the invariance defect (\ref{inv.defect}) and the vector field of (\ref{d}), respectively. $\mbox{\boldmath$\Delta$}$ is evaluated by using the thermodynamic projector (\ref{thermo.projector}). By averaging over all the invariance defect measures, the mean invariance defect is provided: results for both cases are condensed in Table \ref{table01}. Note that the adopted invariance defect measure is dimensionless as it compares the invariance defect with the vector field. Calculations prove that the GQEG is better than the SQEG (straightforwardly extended); nevertheless the SEGQEG construction, since it requires a much lower computational effort and still has an error similar to the GQEG, is recommended when the SQEG is considered not satisfactory (e.g. big mean defect).   
\begin{table}
\centering
\begin{tabular}{l|c|r|}
\textit{}& {1.case} & {2.case}\\
\hline
{SQEG} & {0.318} & {0.645}\\
\hline
{GQEG} & {0.238} & {0.460}\\
\hline
{SEGQEG} & {0.303} & {0.491}\\
\hline
\end{tabular} 
\caption{Mean invariance defect (dimensionless): three approximations of the invariant grid under comparison for the hydrogen oxidation reaction. In 1.case, the parameter set is: $k_1^+=2$, $k_2^+=1$, $k_3^+=1$, $k_4^+=10^3$, $k_5^+=10^3$, $k_6^+=10^2$, $\varepsilon ^2  = 0.45 \cdot 10^3$. In 2.case, the parameter set is: $k_1^+=20$, $k_2^+=1$, $k_3^+=1$, $k_4^+=10^3$, $k_5^+=10^3$, $k_6^+=10^2$, $\varepsilon ^2  = 0.45 \cdot 10^3$.} \label{table01}
\end{table}

\section{Conclusions}\label{conclus}
In this paper, the problem of Quasi Equilibrium Manifold approximation by means of a grid description is addressed. To this end, the notion of Quasi Equilibrium Grid (QEG) is introduced and a proper algorithm to construct it, in any dimension, is suggested (QEGA). It has been shown, through illustrative examples, that the QEGA gives a very good QEM approximation without facing the analytical difficulties of Lagrange multipliers method implementation in large dimension. Since the QEGA is a completely numerical procedure, it reveals particularly suitable for providing the MIG procedure with the first SIM approximation. As it has been illustrated, some proper hybrid procedures $QEGA$ + $MIG$, where both methods are alternatively used, allow to obtain accurate SIM approximations. 
It was proved that two special $QEGA$ + $MIG$ procedures deliver enhanced approximations of SIM: the \textit{Guided Quasi Equilibrium Grid} and the \textit{Symmetric Entropic Guided Quasi Equilibrium Grid}. Here, we want to stress the two major advantages  of the method proposed. First of all, it is a completely numerical algorithm which only deals with nodes sets. Moreover, it is a local construction: namely, the computation of a new node $\mbox{\boldmath$c$}_{n + 1}$, which has to be added to the grid, only depends on the previous neighbor $\mbox{\boldmath$c$}_n$. Those two points make the QEG construction suitable for numerical applications and parallel realizations. Finally, it is not excluded that the QEGA is applicable not only for model reduction, but in some very different fields, too. Indeed, it was mentioned that the QEM notion already is exploited for some applications in Lattice Boltzmann schemes simulations. More generally, the QEGA is a numerical tool which can be used to find a grid-based approximation for the locus of minima of a convex function under some linear constraints. In this paper we focused only on the geometry of the model reduction, that is, construction of slow invariant manifolds approximations. The implementation of grid-based integrators for dynamic equations will be presented in a separate publication. 

\section{Acknowledgments}\label{Acknow}
Prof. A. N. Gorban is gratefully acknowledged for the fruitful discussions about the concept of Quasi Equilibrium Manifold and for suggesting the reaction mechanism (\ref{reaction_1D}). We thank Prof. K. B. Bouluochos, Dr. C. E. Frouzakis for several discussions and suggestions. This work was partially supported by SNF, Project 200021-107885/1 (E.C.) and by BFE, Project 100862 (I.V.K.).

\end{document}